\documentclass[varenna]{cimento}
\usepackage{graphicx,amsmath,amssymb}

\title{Cosmic rays from star clusters}
\author{S.~Gabici\thanks{gabici@apc.in2p3.fr}}
\institute{Universit\'e Paris Cit\'e, CNRS, Astroparticule et Cosmologie, F-75013 Paris, France}

\begin{document}

\maketitle

\begin{abstract}
Massive stars blow powerful winds and eventually explode as supernovae.
By doing so, they inject energy and momentum in the circumstellar medium, which is pushed away from the star and piles up to form a dense and expanding shell of gas.
The effect is larger when many massive stars are grouped together in bound clusters or associations.
Large cavities form around clusters as a result of the stellar feedback on the ambient medium.
They are called superbubbles and are characterised by the presence of turbulent and supersonic gas motions.
This makes star clusters ideal environments for particle acceleration, and potential contributors to the observed Galactic cosmic ray intensity.
\end{abstract}

\section{Introduction}

More than one century after their discovery, revealing the origin of cosmic rays (CRs) remains one of the central open issues in high energy astrophysics.
CRs are energetic particles that hit the Earth atmosphere from outer space.
They are mainly atomic nuclei (mostly protons, with a $\sim$~10\% contribution from helium and $\sim$~1\% of heavier nuclei) plus a contribution from electrons at the percent level \cite{CRbooks}.
Except for the (very few) highest energy particles, CRs are accelerated within the Milky Way, which therefore must host efficient and powerful particle accelerators.

Any scenario proposed to explain the origin of Galactic CRs must satisfy (at least!) the following conditions, inferred from direct and indirect observations of cosmic particles (see e.g. Blasi's lecture notes in this volume):
\begin{enumerate}
\item{sources must inject CRs in the interstellar medium (ISM) at a rate of $\sim 10^{41}$~erg/s \cite{CRpower};}
\item{the energy spectrum of the CRs injected in the ISM must be close to a power law $\propto E^{-s}$ with $s \sim 2.1...2.4$ \cite{andyreview};}
\item{CR protons must be accelerated up to energies exceeding those of the CR knee, a steepening observed in the CR spectrum at a particle energy of few PeV \cite{etienne};}
\item{the observed differences between the composition of CRs and of cosmic (solar) matter must be explained (for a review on CR composition see \cite{composition} or \cite{vincent} and references therein).}
\end{enumerate}

As seen in many of the Chapters in this book, the most common working hypothesis is that Galactic CRs are accelerated at supernova remnant (SNR) shocks via diffusive acceleration \cite{luke,blasireview,lukerecent}.
The main argument in favour of this hypothesis is that the rate at which mechanical energy is injected in the ISM by supernova explosions is $\approx 10^{42}$~erg/s.
Therefore, the observed intensity of CRs can be explained if $\sim$~10\% of such mechanical energy is somehow converted into accelerated particles (see point 1 in the list above). 
Moreover, even though the test-particle theory of diffusive shock acceleration predicts power law spectra of accelerated particles of slope $s = 2$, various kind of non-linearities in the acceleration mechanism can be invoked to explain the required steeper spectra (point 2 in the list) \cite{drift}.

On the other hand, points 3 and 4 above are more difficult to be accounted for (see \cite{myreview} for an extended critical review of the SNR paradigm).
The acceleration of protons beyond PeV energies at SNR shocks requires very large shock velocities and large values of the magnetic field strength.
These conditions might be achieved in the very early stages of the SNR lifetime (during the first few tens of years \cite{schure1}), but it is not clear if in such a short time {\it enough} multi-PeV protons can be produced to match observations \cite{schure2}.

Another major difficulty encountered by the SNR scenario is the explanation of some anomalous isotopic ratios observed in CRs.
Most notably, the $^{22}$Ne/$^{20}$Ne ratio in CRs is a factor of $\approx 5$ larger than the value found in Solar abundances \cite{22Ne}.
This discrepancy can be explained if ejecta from Wolf-Rayet stars, which are enriched in $^{22}$Ne, are accelerated and contribute to the observed CR intensity \cite{casse}.
There are two ways to do so: either Wolf-Rayet material is accelerated at the stellar wind termination shock (WTS) \cite{casse,cesarsky,vincent} or Wolf-Rayet stellar winds pollute the ISM medium with $^{22}$Ne, which is then accelerated by SNR shocks \cite{higdon}.
In both scenarios, star clusters are likely to play a prime role, as massive stars do not form isolated, but rather in groups.

The study of particle acceleration in and around stellar clusters is therefore of great interest.
During the first few million years of the lifetime of a star cluster, stellar winds dominate the mechanical energy output of the systems, and then, when the most massive stars begin to explode, supernovae take over.
As a result of the combined effect of stellar winds and supernova explosions, large cavities are inflated around star clusters \cite{SB,thibaultPhD}.
Such cavities are called {\it superbubbles}, and have been proposed as sites of particle acceleration alternative to SNRs.
The interior of superbubbles is filled by an hot, tenuous, and most likely very turbulent medium.
Turbulence may be generated by the mutual interactions of stellar ejecta, which can be either continuous winds or supernova explosions (e.g. \cite{thibaultPhD} and references therein)
Given these peculiar conditions, it is not clear if CR production in star clusters is simply the sum of the acceleration at recurrent SNR shocks \cite{SBtheoryLingenfelter} or if a different acceleration mechanism has to be invoked \cite{SBtheoryBykov,etienne1,thibault}.
In both cases, the acceleration of particles at stellar wind termination shocks provides an additional contribution to the CR content of these objects (e.g. \cite{thibault}).

The interest in star clusters as particle accelerators was recently revived by the recent detection of a diffuse gamma-ray emission surrounding a number of such objects \cite{felixmassivestars}.
Further detections in both the GeV and TeV (and possibly multi-TeV) gamma-ray domain were reported (e.g. \cite{clustersgamma}).
Such emission proves unambiguously that star clusters can accelerate particles beyond TeV energies.

The goal of these lecture notes is to provide the basic ingredients to understand the mechanisms which are likely responsible for the acceleration of CRs in and around star clusters.
The remaining of the Chapter is structured as follows.
The formation and evolution of an interstellar bubble inflated by a single massive star will be described in Section~\ref{sec:stellarwind}, while the case of a superbubble inflated by the ensemble of stars that form a cluster will be treated in Section~\ref{sec:clusterwind}.
Some basic concepts on particle acceleration in astrophysical environments are given in Section~\ref{sec:hillas}.
The remainder of Section~\ref{sec:acceleration} will be devoted to a description of possible mechanisms for particle acceleration in/around star clusters, operating both at the WTS \ref{sec:accWTS} and in the turbulent superbubble inflated around the cluster \ref{sec:accSB}.
Open problems in the field will be briefly reviewed in Section~\ref{sec:conclusions}.

\section{Interstellar bubble inflated by a massive star wind}
\label{sec:stellarwind}

The first part of this lecture notes provides a description of how the combined effect of stellar winds and supernova explosions affects the medium surrounding a star cluster.
However, it is convenient to consider first the case of an isolated early-type star located in a uniform and pressureless (cold) ISM of mass density $\varrho_0$ \cite{weaver}.
At some time $t = 0$ the star begins to emit a spherically symmetric and steady wind characterised by a mass loss rate $\dot{M}_w$ and by a constant terminal velocity $u_w$.
The wind kinetic power is then $L_w = \dot{M}_w u_w^2 /2$ and its density profile can be derived from mass conservation [$\nabla (\varrho_w u_w) = 0$], to give:
\begin{equation} 
\label{eq:rhow}
\varrho_w = \frac{\dot{M}_w}{4 \pi u_w R^2} ~.
\end{equation}
Here, $R$ represents the distance from the star, which is treated as a point-like source of mechanical energy.

In order to quantify the impact of stellar winds on the ambient ISM, let us recall that they are launched as a result of the transfer of momentum from the stellar photons to matter.
This happens through the absorption and scattering of UV lines \cite{winds}. 
As the luminosity of a star $L_*$ increases steeply with its mass, the most powerful winds are found around the most massive stars.
Using the observed correlation between the momentum carried by the wind and that carried by stellar photons, $\dot{M}_w u_w \approx (1/2) L_*/c$, the kinetic power of the wind can be expressed as \cite{krumholzbook}:
\begin{equation}
L_w \approx \frac{L_* u_w}{4~c} \sim 3 \times 10^{36} \left( \frac{L_*}{3 \times 10^5 L_{\odot}} \right) \left( \frac{u_w}{3000~{\rm km/s}} \right) ~ \rm erg/s
\end{equation}
where $L_{\odot}$ is the luminosity of the Sun, and where quantities have been normalised to typical values of very massive stars (several tens of solar masses).
At this point, we can estimate the total energy output integrated on the lifetime of the star $\tau_*$ (for very massive stars this is of the order of few million years) to get:
\begin{equation}
E_w = L_w \tau_* \sim 4 \times 10^{50} \left( \frac{L_*}{3 \times 10^5 L_{\odot}} \right) \left( \frac{u_w}{3000~{\rm km/s}} \right) \left( \frac{\tau_*}{4~{\rm Myr}} \right) ~ \rm erg
\end{equation}
Remarkably, this is of the same order as the energy deposited in the ISM by a supernova explosion ($\approx 10^{51}$ erg), and therefore stellar winds from very massive stars are expected to impact dramatically on the ambient ISM.

In particular, as a result of the injection of mechanical energy, cavities are inflated in the ISM around massive stars.
Such cavities are called {\it interstellar bubbles}, and their evolution in time proceeds through a number of different phases, which will be described in the following.

\subsection{Free-expansion phase}

At first, the circumstellar matter is pushed away by the wind and accumulates in a dense, expanding, and spherical shell located at a distance $R_s$ from the star.
Initially, the shell of swept up interstellar gas contains very little mass, and therefore the wind expands freely ($R_s \sim u_w t$, hence the name {\it free expansion phase}).
As the shell moves at a highly supersonic velocity, a shock wave, called {\it forward shock}, forms ahead of it.
Then, when the mass of interstellar gas swept up by the shock, $M_{sw} = (4 \pi/ 3) \varrho_0 R_s^3$, becomes comparable to the mass carried by the wind, $\dot{M}_w t$, the inertia of the shell becomes important and the expansion decelerates.
This happens at a time:
\begin{equation}
\tau_f = \left( \frac{3 \dot{M}_w}{4 \pi \varrho_0 u_w^3} \right)^{\frac{1}{2}} \sim 16 \left( \frac{\dot{M}_w}{10^{-6} M_{\odot}/{\rm yr}}\right)^{\frac{1}{2}} \left( \frac{n_0}{{\rm cm}^{-3}}\right)^{-\frac{1}{2}} \left( \frac{u_w}{3000~{\rm km/s}} \right)^{-\frac{3}{2}} \rm yr
\end{equation}
where the mass loss rate has been normalised to a value appropriate to describe a main sequence star of several tens of solar masses \cite{winds}, and the ambient gas number density $n_0 = \varrho/\mu m_H$ to a value characteristic of the interstellar gas \cite{ISM}.
Here, $m_H$ is the mass of hydrogen, and $\mu \sim 1.4$ accounts for the presence of helium in the ISM at the $\sim$~10\% level.
Note that the phase of free expansion is several orders of magnitude shorter than the lifetime of a massive star ($\tau_*$, few million years), and therefore will not be further discussed in the following.

\subsection{Adiabatic phase}

\begin{figure}
\centering
\includegraphics[width=0.7\textwidth]{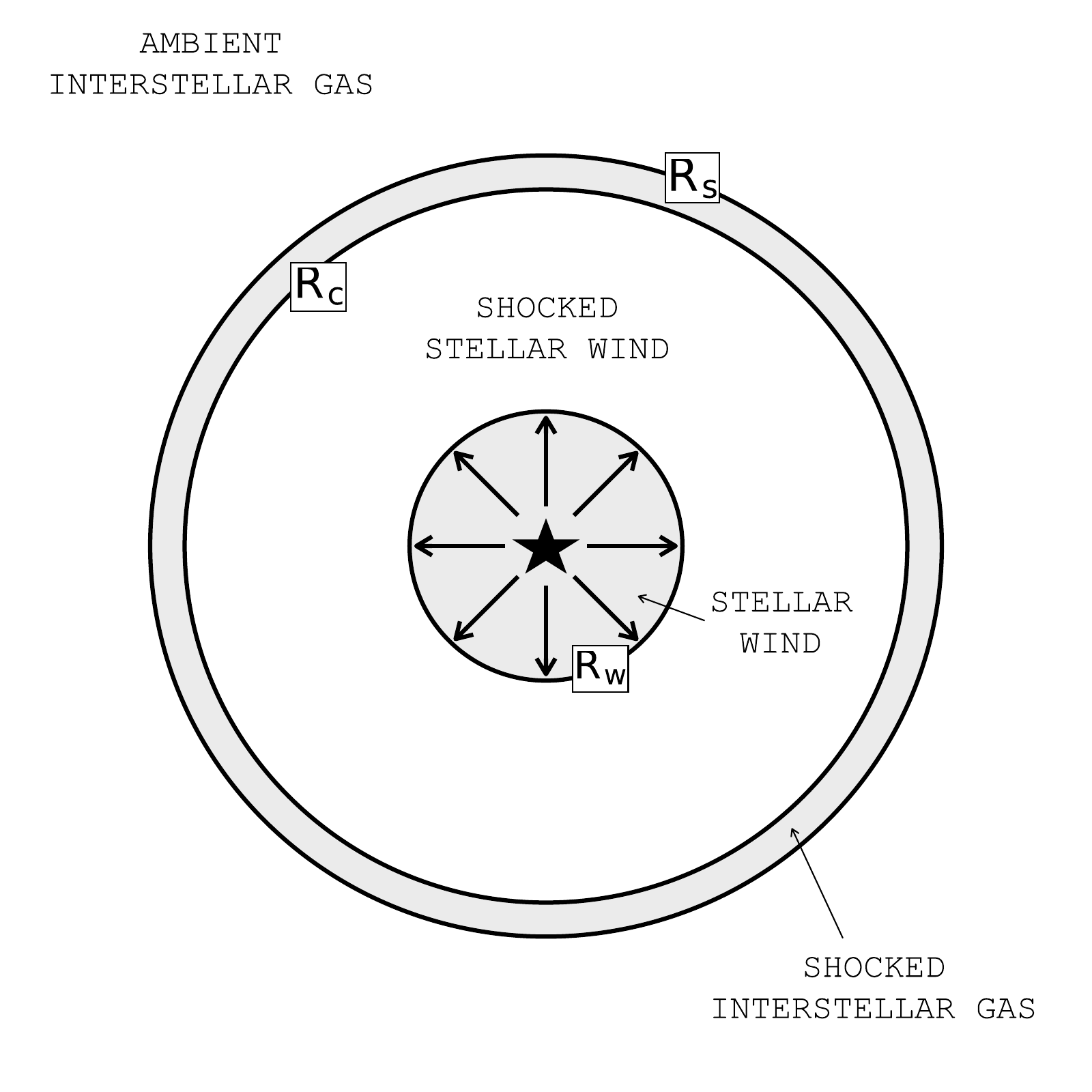}     
\caption{Structure of an interstellar bubble inflated by a massive star wind. See text for details.}
\label{fig:bubble}
\end{figure}

After the free expansion phase, the shell begins to decelerate, and therefore the wind no longer expands freely.
The deceleration of the wind takes place at a spherical shock wave, called {\it wind termination shock}.
The resulting structure is called {\it interstellar bubble} and consists of four regions (see Fig.~\ref{fig:bubble}).
Proceeding from the star outwards they are: {\it i)} an highly supersonic wind; {\it ii)} a region containing the shocked wind material; {\it iii)} a shell containing the shocked interstellar gas; {\it iv)} the ambient ISM.
Regions {\it i} and {\it ii} are separated by the WTS, located at $R = R_w$, regions {\it ii} and {\it iii} by a contact discontinuity ($R = R_c$), and regions {\it iii} and {\it iv} by the forward shock ($R = R_s$).

At this point, it is useful to estimate the thickness of the shell of shocked interstellar gas, $\Delta R = R_s-R_c$.
As the forward shock, at least in the early phase of the expansion, is certainly very strong\footnote{This implies that the assumption of a cold ambient interstellar gas is appropriate. It can be seen by recalling that the flux of momentum crossing a shock which moves at velocity $u_s$ is $\varrho_0 u_s^2 + P_0$, where $\varrho_0 u_s^2$ is the shock ram pressure and $P_0$ is the pressure of the ISM upstream of the shock. For a strong shock $u_s \gg c_s$ and therefore $P_0 \sim \varrho_0 c_s^2 \ll \varrho_0 u_s^2$ is much smaller than the ram pressure and therefore can be neglected.} (the sound speed in the warm ISM is $\approx 10$~km/s), one can safely assume that the density of the gas in the shell is that of the ambient ISM compressed by a factor of 4 (see Caprioli's lecture notes in this volume or \cite{landau}).
The mass of gas in the shell can be computed as $M_{sh} \sim (4 \pi R_s^2 \Delta R) (4 \varrho_0)$, and it must be equal to the total mass of the shocked ISM, $M_{sh} = (4 \pi/3) R_s^3 \varrho_0$.
Equating the two definitions of $M_{sh}$ gives $\Delta R \sim 0.08 R_s$, which means that the shell is quite thin.
Therefore, in order to simplify the problem, the shell will be assumed to be infinitely thin ($R_s \equiv R_c$), which is, the position of the shell coincides with that of the forward shock.
While this might seem to be a rather crude approximation, it provides in fact reasonably accurate results. 

As long as the system is adiabatic (i.e. radiative losses can be neglected), the expansion rate of the forward shock can be derived in a very simple way using dimensional analysis. 
This can be done because the wind kinetic power, $L_w$, is dissipated at the WTS and mostly converted into internal energy (and pressure) of the gas in region {\it ii}.
The pressure of the gas in that region pushes onto the shell (region {\it iii}), whose inertia depends on the density of the ambient medium $\varrho_0$.
It follows that the expansion rate of the forward shock must depend uniquely on the values of $L_w$ and $\varrho_0$.
As it is not possible to combine these two quantities to obtain a characteristic spatial or temporal scale of the problem, the solution has to be scale free, i.e., a power law: $R_s \propto t^{\alpha}$. 
The only possible scale-free solution is then: 
\begin{equation}
\label{eq:ss}
R_s \sim a \left( \frac{L_w}{\varrho_0} \right)^{1/5} t^{3/5} 
\end{equation}
where $a$ is a non-dimensional constant of order unity.
The expansion velocity of the shell is given by: 
\begin{equation}
\label{eq:uad}
u_s = \frac{{\rm d}R_s}{{\rm d}t} = \frac{3}{5} \frac{R_s}{t} = \frac{3}{5} a \left( \frac{L_w}{\varrho_0} \right)^{1/5} t^{-2/5}
\end{equation}
For a rigorous discussion on scale-free (or self-similar) solutions the reader is referred to \cite{zeldovich}.
Note that the shocked ambient gas will be heated up to very large temperatures. 
Behind a strong shock the temperature of the gas is (see Caprioli's lectures, this volume, or \cite{landau}):
\begin{equation}
\label{eq:kT}
kT = \frac{3}{16} \mu m_H u_s^2  \sim 44 ~a^2 \left( \frac{L_w}{10^{36} {\rm erg/s}} \right)^{2/5} \left( \frac{n_0}{{\rm cm}^{-3}} \right)^{-2/5} \left( \frac{t}{10^4~{\rm yr}} \right)^{-4/5} ~ \rm eV
\end{equation}
where $k$ is the Boltzmann constant, and Eq.~\ref{eq:uad} was used to compute the second equality.
A plasma characterised by such temperatures radiates in the UV/soft-X ray domain and cools in a characteristic time $\tau_c$, which mostly depends on the gas temperature and density. Therefore, the system evolves in the {\it adiabatic phase} for $\tau_f < t < \tau_{ad} \approx \tau_c$.

Radiative losses are  conveniently described by a cooling function $\Lambda(T)$ (erg cm$^3$/s) which depends on gas temperature and metallicity (here assumed to be solar).
For the hot and ionised plasmas considered here, the cooling is dominated by both line and continuum thermal emission.
In the range of temperatures $10^{5}~{\rm K} \lesssim T \lesssim 10^{7.5}$~K the cooling function can be (roughly) approximated as $\Lambda(T) = \Lambda_0 T^{-1/2} \sim 1.6 \times 10^{-19} T^{-1/2}$~erg cm$^3$/s \cite{cioffi}.
A fully ionised plasma characterised by an hydrogen number density $n_H$ and an electron number density $n_e \sim 1.2 n_H$ (the numerical factor accounts for the presence of helium) cools at a rate $L \equiv n_H n_e \Lambda(T) \sim 1.2 ~n_H^2 \Lambda(T)$.
Due to the $\propto n_H^2$ scaling, the shell of shocked ambient gas cools first, as it is the densest region in the system (it contains most of the total mass concentrated in a very small volume).
As the thermal energy density of a fully ionised plasma is $\epsilon_{th} \sim 2.3 (3/2) n_H k T$, where $k$ is the Boltzmann constant, the cooling time of the plasma in the shell can be written as:
\begin{equation}
\tau_c = \frac{\epsilon_{th}}{L}  = \frac{2.3 (k T)^{3/2}}{3.2~ n_0 \Lambda_0 k^{1/2}} \sim 2.5 \times 10^4 \left( \frac{n_0}{{\rm cm}^{-3}} \right)^{-1} \left( \frac{kT}{0.1~{\rm keV}} \right)^{3/2}  \rm yr
\end{equation}
where $n_H = 4 \times n_0$ to account for shock compression.

The cooling time of the shell can be now estimated by equating $\tau_c$ to the age of the system $t$.
When that is done (using Eq.~\ref{eq:kT}) one gets a duration of the adiabatic phase equal to:
\begin{equation}
t_{ad} \sim 8.6 \times 10^3 ~ a^{15/11} \left( \frac{L_w}{10^{36} {\rm erg/s}} \right)^{3/11} \left( \frac{n_0}{{\rm cm}^{-3}} \right)^{-8/11} \rm yr
\end{equation}
which is much smaller than the lifetime of the system $\tau_*$.
For this reason, the adiabatic phase will not be further discussed.

\subsection{Partially radiative, or snowplow phase}
\label{sec:snowplow}

After $t_{ad}$, then, the shell cools but the material injected by the wind into zone {\it ii} is still adiabatic.
The density in region {\it ii} will be shown to be orders of magnitudes smaller than the density in the shell, and therefore the interior will cool much later.
It follows that interstellar bubbles spend most of their life in this partially radiative phase, which deserves to be studied in detail.

Remarkably, Eq.~\ref{eq:ss} provides a good description of the expansion rate of the forward shock also in this phase.
Calculations more accurate than those performed here show that the only difference is that the value of the constant $a$ is equal to 0.88 in the fully adiabatic phase, and decreases to 0.76 when the shell becomes radiative \cite{weaver}.
In order to understand why this is the case, assume that all the kinetic energy that flows across the forward shock is radiated away.
The rate at which the system loses energy is then:
\begin{equation}
\label{eq:Lrad}
L_{rad} = \left( 4 \pi R_2^2 \right) \left( \frac{1}{2} \varrho_0 u_s^3 \right) 
\end{equation}
Such a rate is constant in time if the scalings $R_s \propto t^{3/5}$ and $u_s \propto t^{-2/5}$ are adopted.
This means that it is possible to define an effective injected power as $L_{eff} = L_w - L_{rad}$, which is also constant in time.
Thus, Eq.~\ref{eq:ss} and \ref{eq:uad} are still solutions of the problem after the substitution $L_w \rightarrow L_{eff}$.
After setting $a = 0.76$ one finally gets:
\begin{eqnarray}
\label{eq:Rs}
R_s &\sim& 26 \left( \frac{L_w}{10^{36}{\rm erg/s}} \right)^{1/5} \left( \frac{n_0}{{\rm cm}^{-3}} \right)^{-1/5} \left( \frac{t}{{\rm Myr}} \right)^{3/5} \rm pc \\
u_s &\sim& 15 \left( \frac{L_w}{10^{36}{\rm erg/s}} \right)^{1/5} \left( \frac{n_0}{{\rm cm}^{-3}} \right)^{-1/5} \left( \frac{t}{{\rm Myr}} \right)^{-2/5} \rm km/s
\label{eq:us}
\end{eqnarray}

The total energy in the system at a time $t$ can be computed from Eq.~\ref{eq:Lrad}, \ref{eq:Rs}, and \ref{eq:us} as:
\begin{equation}
\label{eq:Etot}
E_{tot} \sim (L_w-L_{rad}) t = \left[ 1 - 2 \pi \left( \frac{3}{5} \right)^3 a^5 \right] L_w t
\end{equation}
Recalling that during this phase the system is composed by a cold and dense expanding shell, pushed by an hot and rarefied interior, the total energy can be written as the sum of the kinetic energy of the shell:
\begin{equation}
\label{eq:Ek}
E_k = \frac{1}{2} M_{sh} u_s^2 = \frac{2 \pi}{3} \left( \frac{3}{5} \right)^2 a^5 L_w t 
\end{equation}
plus the thermal energy of the hot interior:
\begin{equation}
\label{eq:Eth}
E_{th} = \left( \frac{3}{2} P \right) \left( \frac{4 \pi}{3} R_s^3 \right)
\end{equation}
where $P$ is the average gas pressure in region {\it ii}.
Combining Eq.~\ref{eq:Etot}, \ref{eq:Ek}, and \ref{eq:Eth} one can see that $E_k/E_{tot} \sim 0.3$ and $E_{th}/E_{tot} \sim 0.7$ and that the pressure in the hot interior is \cite{weaver}:
\begin{equation}
\label{eq:P}
P = 4.5 \times 10^{-12} \left( \frac{L_w}{10^{36}{\rm erg/s}} \right)^{2/5} \left( \frac{n_0}{{\rm cm}^{-3}} \right)^{3/5} \left( \frac{t}{{\rm Myr}} \right)^{-4/5} ~ \rm erg/cm^3
\end{equation}

The expressions above for $R_s$, $u_s$, and $P$ have been derived under the assumption of a cold (pressureless) ambient medium.
Such assumption is valid as long as the forward shock is strong.
The shock Mach number is obtained dividing the shock velocity by the sound speed of the ISM of temperature $T_0$, $c_{s,0} = \sqrt{(5/3) kT_0/\mu m_H}$, which gives:
\begin{equation}
{\cal M}_s \sim 1.5 \left( \frac{L_w}{10^{36}{\rm erg/s}} \right)^{1/5} \left( \frac{n_0}{{\rm cm}^{-3}} \right)^{-1/5} \left( \frac{T_0}{10^4~{\rm K}} \right)^{-1/2} \left( \frac{t}{{\rm Myr}} \right)^{-2/5}
\end{equation}
This shows that, for a warm ISM characterised by a temperature of $T_0 \sim 10^4$~K, Eq.~\ref{eq:Rs}, \ref{eq:us}, and \ref{eq:P} are valid only up to $t_s \lesssim 1$~Myr.
After that, the pressure of the ambient medium starts to be important, and the expansion rate of the shell drops significantly: the bubble enters the {\it pressure-confined} phase \cite{weaver,koo}.

\subsubsection{\it The internal structure of interstellar bubbles}

Although the radiative cooling of the shell has little impact on the expansion rate of the forward shock, it strongly affects the internal structure of the system.
First of all, as a consequence of radiative cooling, the shell collapses and becomes extremely thin and dense.
This can be easily seen by considering an {\it isothermal} forward shock, i.e., a shock were radiative losses in the denser downstream region are so effective to cool the gas down to the initial (upstream) temperature \cite{shu}.
If the temperature is constant across the shock transition, the sound speed, which depends on temperature only, will be equal to the interstellar value $c_{s,0}$ on both sides of the shock.
This means that the pressure will depend on density only, $P_i = \varrho_i c_{c,0}^2$, where the subscript refers to the upstream ($i = 1$) or downstream ($i = 2$) region.
The conservation of momentum flux across the shock then reads:
\begin{equation}
\varrho_1 u_1^2 + \varrho_1 c_{s,0}^2 = \varrho_2 u_2^2 + \varrho_2 c_{s,0}^2
\end{equation}
which can be divided by $\varrho_1 u_1^2$ and combined with the condition for mass conservation $\varrho_1 u_1 = \varrho_2 u_2$ to give:
\begin{equation}
\left( r - {\cal M}^2 \right) (r - 1) = 0
\end{equation}
where ${\cal M} = u_1/c_{s,0}$ is the shock Mach number and $r = \varrho_2/\varrho_1 = u_1/u_2$ is the shock compression factor.
Neglecting the solution $r = 1$, which is unphysical (no shock wave), one is left with $r = {\cal M}^2$.
Then, for strong shocks the compression factor can largely exceed 4 and as a consequence the shell becomes very thin (hence the name {\it snowplow phase} as shocked ambient matter accumulates just behind the forward shock).
It follows that the approximation made above of an infinitesimally thin shell is even more appropriate during the partially radiative phase as long as the Mach number ${\cal M}_s$ is significantly large. 

The arbitrarily large compression for an arbitrary large Mach number implied by $r = {\cal M}^2$ is of course not physical.
In fact, also the interstellar magnetic field will be compressed at the shock, as $B_2 \sim r B_1$. 
Such compression induces an increase of the downstream magnetic pressure with the shock compression factor scaling as $P_{B,2} = B_2^2/8 \pi \propto r^2$.
This scaling is steeper than that of the downstream thermal pressure $\varrho_2 u_2^2 \propto r$.
Therefore, for large compression factors, the pressure downstream of the shock is largely dominated by the magnetic one.
For an highly supersonic (the upstream gas pressure can be neglected) and highly superalfvenic (the upstream magnetic pressure can be neglected\footnote{A superalfvenic shock moves at a speed larger than the Alfv\'en one, $u_1 > v_A = B_1/(4 \pi \varrho_1)^{1/2}$. This can be rewritten as $\varrho_1 u_1^2 > B_1^2/4 \pi = 2 P_{B,1}$. Then, for highly superalfvenic shocks ($u_1\gg v_A$) the magnetic pressure upstream is negligible when compared to the ram pressure.}) momentum conservation simplifies to:
\begin{equation}
\varrho_1 u_1^2 \sim r^2 \frac{B_1^2}{8 \pi}
\end{equation}
which implies that the compression factor does not increase indefinitely with the Mach number ${\cal M}$, but is bounded to the value \cite{radiativeshock}:
\begin{equation}
\label{eq:rmax}
r \sim \left( \frac{8 \pi \varrho_1 u_1^2}{B_1^2} \right)^{1/2} = \sqrt{2} {\cal M}_A \gg 1
\end{equation}
where ${\cal M}_A = u_1/v_A$ is the {\it alfvenic} Mach number.
Eq.~\ref{eq:rmax} shows that for a strong and magnetised shock the compression factor can still be very large, but never diverges.

The thin, cold, and magnetised shell of swept up ISM bounds the low density cavity, which is filled with shocked wind material and is therefore hot.
This has two consequences. 
First, a hot gas is characterised by a large speed of sound.
Under these conditions sound waves can cross the cavity in a time which is shorter than the age of the system.
Therefore, the pressure $P$ in region {\it ii} can be assumed to be (roughly) spatially uniform. 
Second, thermal conduction will operate at the interface between the cold shell and the hot interior, causing cold gas to evaporate from the shell into the cavity and mix with the shocked wind material \cite{spitzer}.

Due to thermal conduction, then, the boundary between region {\it ii} and {\it iii} is not sharp, but it is smeared out.
It is convenient to describe the transition region in the rest frame where the inner boundary of the shell is at rest, and to assume that the inward flow of evaporating material is well described by a stationary one dimensional (plane-parallel) isobaric flow.
If radiative losses are assumed to be unimportant in the transition region, and if the role of the magnetic field is ignored, the gas flow is obtained after balancing the outward heat flux due to thermal conduction with the inward mechanical energy flow carried by the evaporating gas.

Heat flow from region {\it ii} to region {\it iii} is proportional to the temperature difference between the two regions, and can be written as:
\begin{equation}
F_h = -K \frac{\partial T}{\partial z} 
\end{equation}
where $z$ is the distance from the shell and the minus sign indicates that heat flows towards the colder region. 
The proportionality coefficient $K$ is called thermal conductivity and depends quite strongly on the gas temperature: $K = C~ T^{5/2}$ \cite{spitzer}. 
On the other hand, $C$ depends weakly on temperature (through the Coulomb logarithm) and will be therefore treated as a constant: $ C = 1.2 \times 10^{-6}$~erg/cm/s/K$^{7/2}$ \cite{spitzer,weaver}.
If radiative losses in the transition region are neglected, at equilibrium the heat flow has to be balanced by a mechanical flow in the opposite direction, that can be estimated as:
\begin{equation}
F_m = \frac{5}{2} P v
\end{equation}
where $v$ is the flow speed and $(5/2) P$ is the specific enthalpy of the gas.
Balancing the flows gives:
\begin{equation}
\label{eq:Ptc}
P \approx \frac{2 ~C ~T^{7/2}}{5~ R_s~ u_s}
\end{equation}
where the crude approximations $\partial/\partial z \approx R_s$ and $v \approx u_s$ were made.

Equating Eq.~\ref{eq:P} and \ref{eq:Ptc} one gets the expression for the time evolution of the temperature in the hot interior:
\begin{equation}
\label{eq:Tb}
T \sim 1.0 \times 10^6 \left( \frac{L_w}{10^{36}{\rm erg/s}} \right)^{8/35} \left( \frac{n_0}{{\rm cm}^{-3}} \right)^{2/35} \left( \frac{t}{{\rm Myr}} \right)^{-6/35} \rm K
\end{equation}
and that for the hydrogen density (as $P \sim 2.3 n k T$):
\begin{equation}
\label{eq:nb}
n \sim 1.3 \times 10^{-2} \left( \frac{L_w}{10^{36}{\rm erg/s}} \right)^{6/35} \left( \frac{n_0}{{\rm cm}^{-3}} \right)^{19/35} \left( \frac{t}{{\rm Myr}} \right)^{-22/35} \rm cm^{-3}
\end{equation}
It should be noted that the contribution from evaporated matter to the total mass inside the bubble is largely dominant when compared to the mass injected by the stellar wind $\dot{M}_w t$. 
This can be easily seen by computing the density one would expect if only shocked wind material were present in the cavity.
Such a density would be: 
\begin{equation}
n \sim \frac{\dot{M}_w t}{\frac{4 \pi}{3} R_s^3} \sim 4 \times 10^{-4} \left( \frac{\dot{M}_w}{10^{-6} M_{\odot}/{\rm yr}}\right) \left( \frac{L_w}{10^{36}{\rm erg/s}} \right)^{-\frac{3}{5}} \left( \frac{n_0}{{\rm cm}^{-3}} \right)^{\frac{3}{5}} \left( \frac{t}{{\rm Myr}} \right)^{-\frac{4}{5}} \rm cm^{-3}
\end{equation}
which is much smaller than the value provided by Eq.~\ref{eq:nb}.

\begin{figure}
\centering
\includegraphics[width=0.7\textwidth]{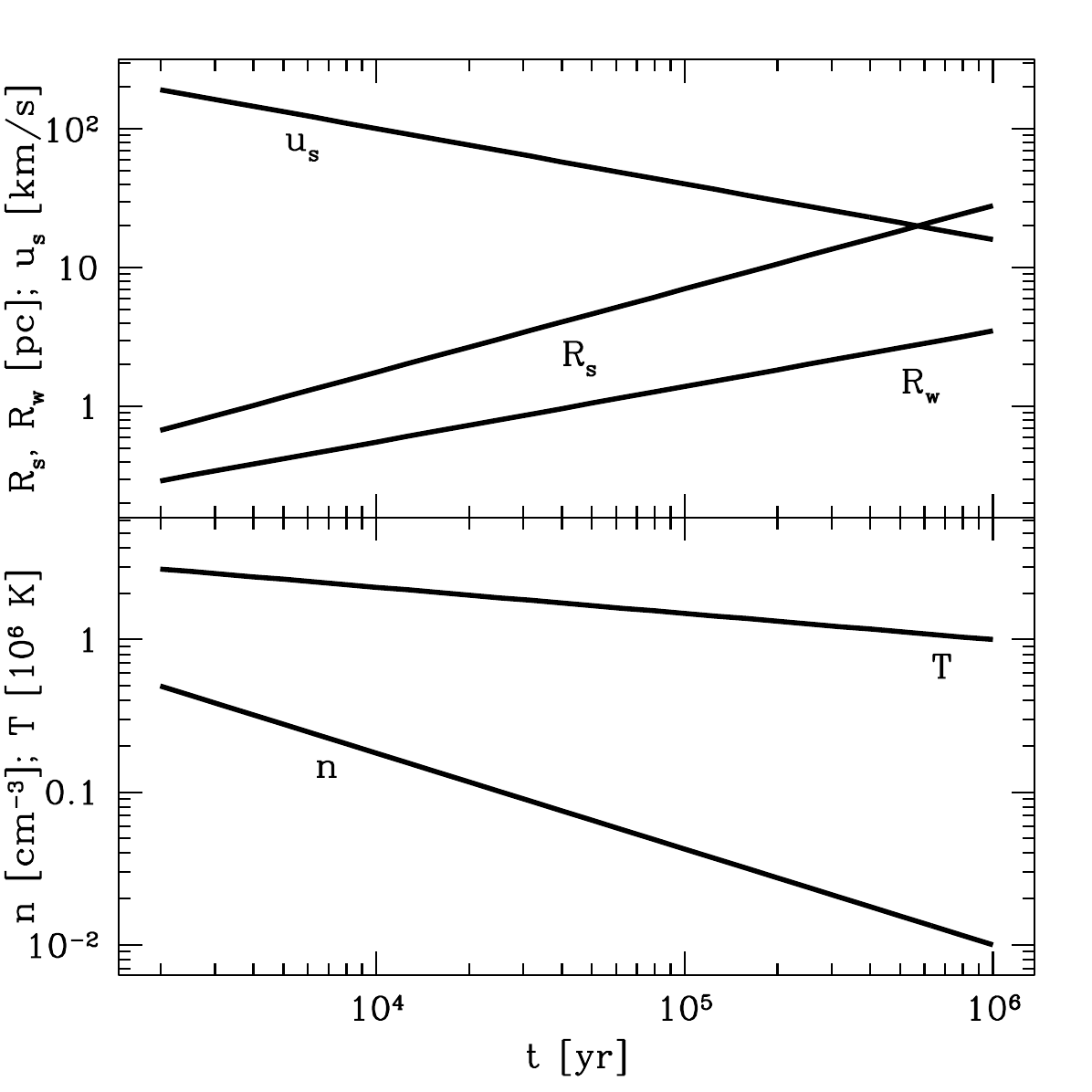}     
\caption{{\bf Top panel:} Time evolution of the forward shock radius $R_s$ and velocity $u_s$ and of the WTS radius $R_w$ for an interstellar bubble inflated in an ISM of density $n_0 = 1$~cm$^{-3}$ by a massive star wind of kinetic power $L_w = 10^{36}$~erg/s and terminal velocity $u_w = 3000$~km/s. {\bf Bottom panel:} time evolution of the temperature $T$ and density $n$ in the interior of the bubble. The time interval on the x-axis spans from the end of the adiabatic phase to the beginning of the pressure-confined phase.}
\label{fig:weaver}
\end{figure}

\subsubsection{The wind termination shock}

Once the internal structure of the bubble has been determined, the only missing piece of information is the evolution in time of the WTS.
An estimate of the position of the shock can be obtained by equating the ram pressure of the wind, $\varrho_w u_w^2$, to the thermal pressure inside the bubble, provided by Eq.~\ref{eq:P}.
By making use of Eq.~\ref{eq:rhow}, this gives \cite{weaver,morlino}:
\begin{equation}
\label{eq:WTS}
R_w \sim 3.5  \left( \frac{L_w}{10^{36}{\rm erg/s}} \right)^{3/10} \left( \frac{n_0}{{\rm cm}^{-3}} \right)^{-3/10} \left( \frac{u_w}{3000~{\rm km/s}} \right)^{-1/2} \left( \frac{t}{{\rm Myr}} \right)^{2/5} \rm pc
\end{equation}
This implies that the WTS expands at a rate which is slower than that of the forward shock, and when the system is well into the snowplow phase the condition $R_w \ll R_s$ is always satisfied.

The main results obtained in this Section are summarised in Fig.~\ref{fig:weaver}, where the evolution in time of the main physical quantities defining an interstellar bubble has been plotted.



 \section{Interstellar bubble inflated by a cluster of massive stars}
 \label{sec:clusterwind}
 
 What happens when a bubble is not inflated by a single star, but rather by a group of them, bundled in a star cluster?
 This situation is indeed very relevant, as most massive stars form in groups or clusters, as the result of the gravitational collapse of dense molecular clouds \cite{krumholzbook}.
 Their short lifetime, combined with a relatively low velocity dispersion, explains why very massive stars are often found in associations.
 This is because they explode as supernovae before having the time to move away from the site of their formation.
 This fact has a very important implication: all the massive stars belonging to a given cluster deposit large amounts of kinetic energy (in form of wind or supernova ejecta) within a small volume.
 
\begin{figure}
\centering
\includegraphics[width=0.45\textwidth]{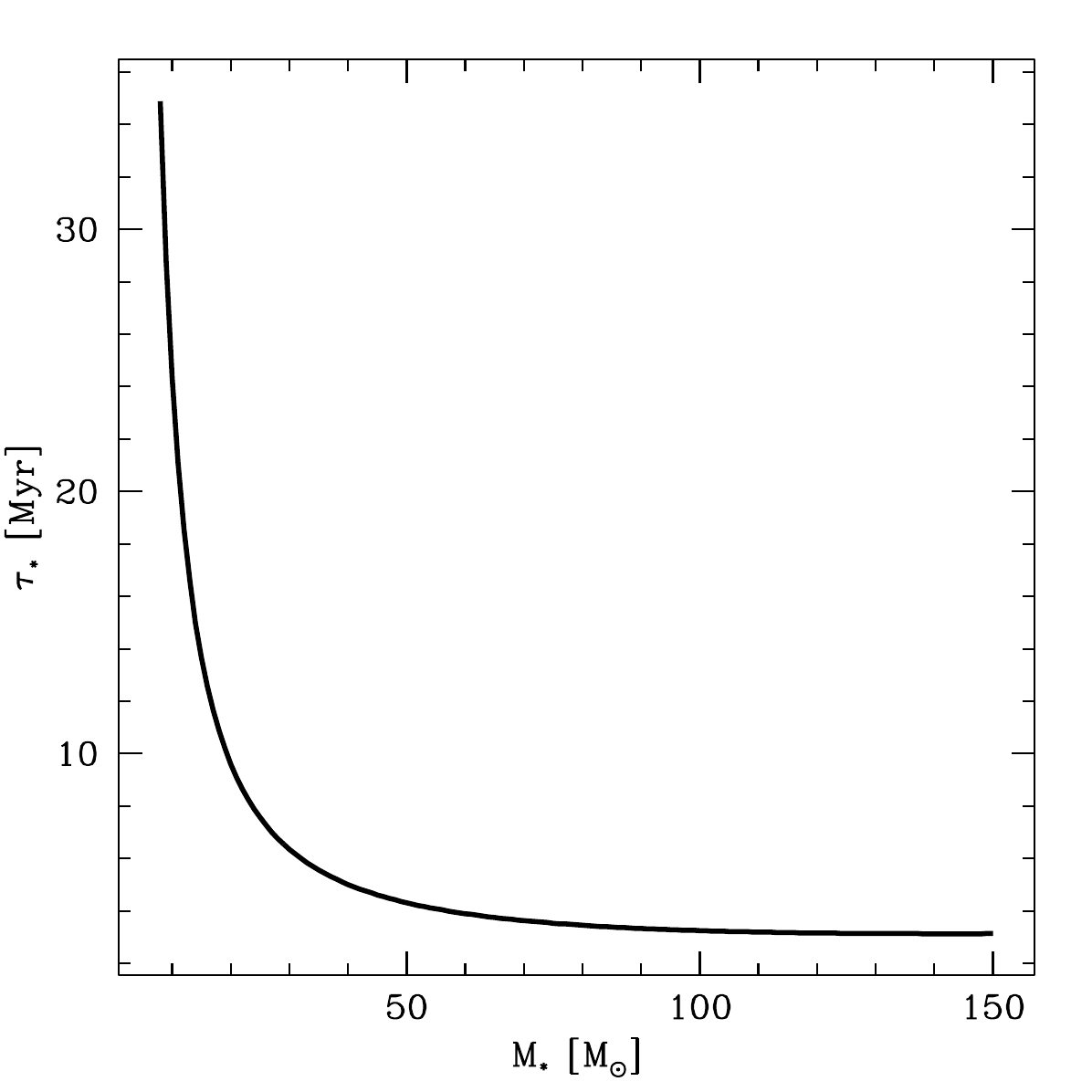}     
\includegraphics[width=0.45\textwidth]{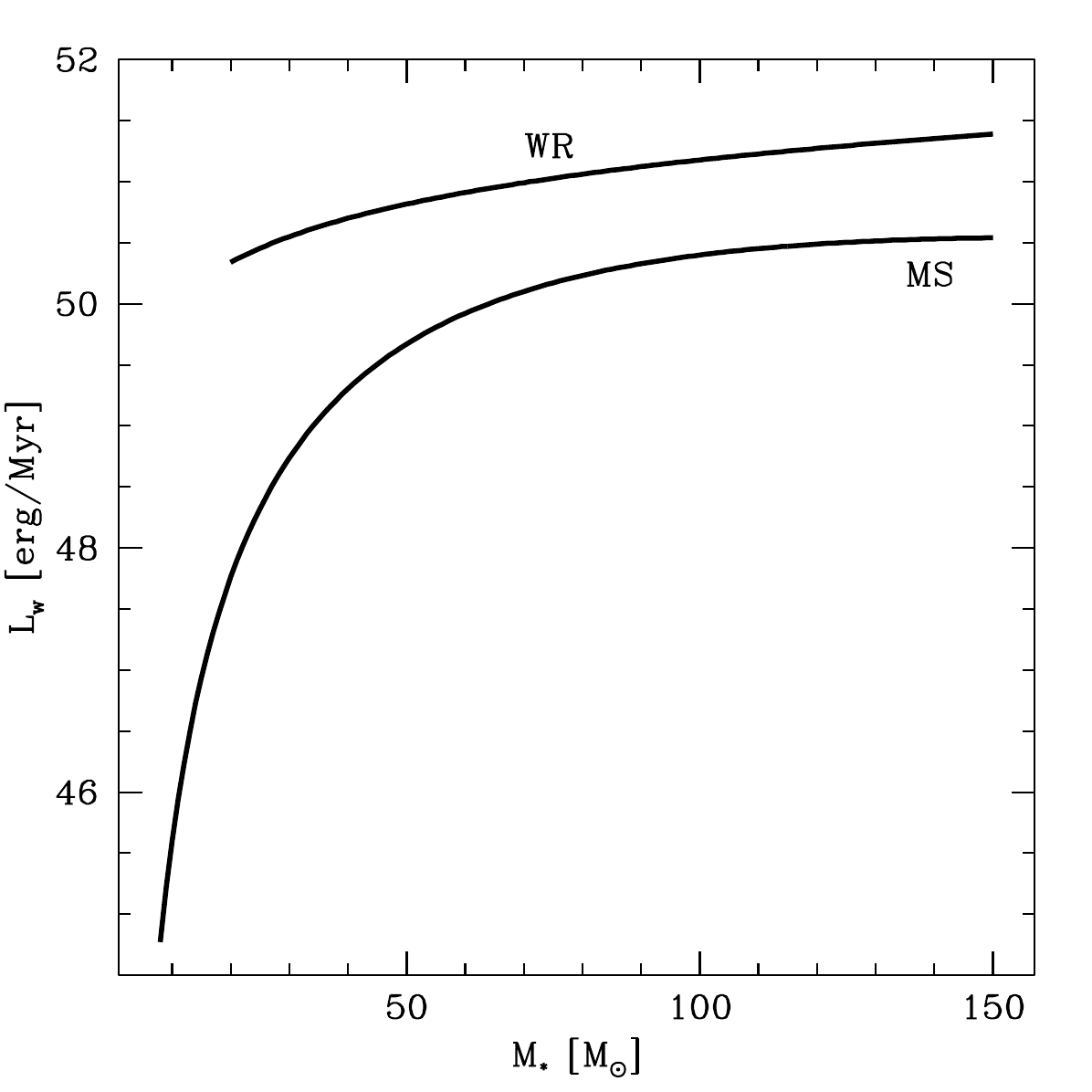}
\caption{{\bf Left panel:} Lifetime of a star as a function of its mass according to the stellar evolution simulations by \cite{limongi}. {\bf Right panel:} Wind power for stars of mass $M_*$ during the main sequence (MS) and Wolf-Rayet (WR) phases \cite{seo,thibault}.}
\label{fig:limongi}
\end{figure}

 Most of the energy is deposited by cluster stars of mass $\gtrsim 10 ~M_{\odot}$.
 These stars emits powerful winds and eventually explode as supernovae.
 They are also characterised by a short lifetime $\tau_*$, which correlates with the initial stellar mass $M_*$, as derived from the stellar evolution model shown in the left panel of Fig.~\ref{fig:limongi}.
 It can be seen from the plot that the star lifetime is a decreasing function of its mass, and spans from few tens of Myr for stars of $\approx 10~M_{\odot}$, down to few Myr for the most massive stars of mass $\approx 100-200~M_{\odot}$
 It follows that, during the first few Myr of the life of a star cluster, stellar winds are the only relevant sources of kinetic energy in the surrounding ambient medium.
 
Stellar evolution models also provide an estimate of the wind power throughout the star life \cite{seo}.
Massive stars spend most of their life in the main sequence, and move to the red supergiant phase at the end of their lives, or to the Wolf-Rayet phase if their mass is large enough ($M_* \gtrsim 20~M_{\odot}$).
Main sequence and Wolf-Rayet winds provide the largest contributions to the total output of kinetic energy.
In particular, the Wolf-Rayet phase lasts for a quite short time, of the order of few times $10^5$~yr, but winds of Wolf-Rayet stars are much more powerful than the main sequence ones, and are likely to dominate the total wind-related kinetic energy output from a star.
The wind power as a function of the initial mass of the star is shown in the right panel of Fig.~\ref{fig:limongi} for both the main sequence and the Wolf-Rayet phase \cite{seo,thibault}.

\begin{figure}
\centering
\includegraphics[width=0.7\textwidth]{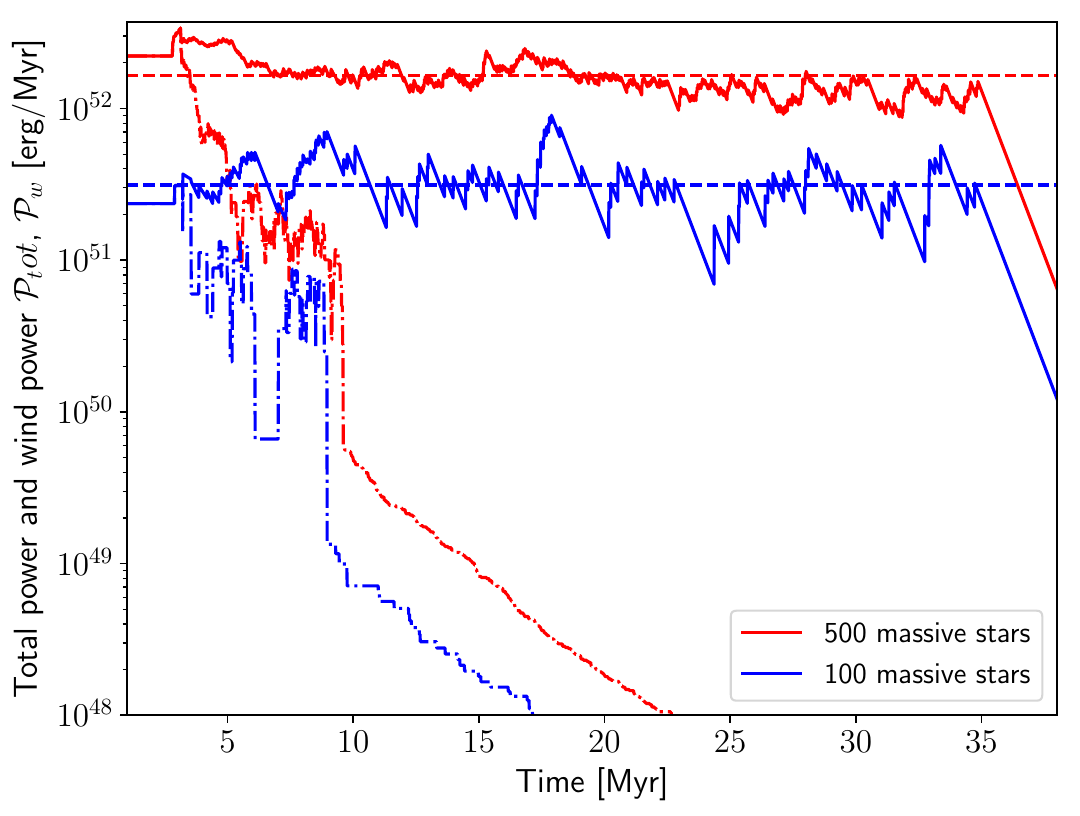}     
\caption{Mechanical power injected by stellar winds (dot-dashed lines) and stellar winds plus supernova explosions (solid lines) as a function of the age of the cluster. The injection of energy at supernova explosions is not instantaneous, but is released over a relaxation time of $\sim 1$~Myr. Red (blue) lines refer to a cluster made of 500 (100) massive stars. Dashed lines show the average cluster power. Figure from \cite{thibault}, where more details can be found.}
\label{fig:luminosity}
\end{figure}

For definiteness, consider a cluster composed of $N_*$ massive stars with masses in the range 8-150~$M_{\odot}$.
The distribution of stellar masses at formation ${\rm d}n_*/{\rm d}M_*$ is called {\it initial mass function} and has been constrained from observations \cite{IMF}.
It is well described by a power law ${\rm d}n_*/{\rm d}M_* = A ~M_*^{-\alpha}$ with slope in the range $\alpha \sim 2.3-2.7$.
The initial mass function can be sampled in order to simulate the masses of all the stars in a cluster.
Then, using the information from Fig.~\ref{fig:limongi}, it is possible to evaluate the cumulaitve mechanical power injected by all stellar winds in the cluster. 
This was done in \cite{thibault}, where it was assumed $\alpha = 2.3$ and that stars with masses larger than 20~$M_{\odot}$ at the end of their life go through a Wolf-Rayet phase lasting 320 kyr.
Results are shown in Fig.~\ref{fig:luminosity} with red and blue dot-dashed lines referring to clusters containing $N_* = 500$ and 100 massive stars, respectively.
The power injected by winds stays roughly constant for the first few Myr of the life of the cluster, and then drops quite quickly as the most massive stars explode as supernovae.
The average power of the winds of massive stars over the entire lifetime of the cluster ($\sim$~35 Myr) is of the order of $\lesssim 10^{35}$~erg/s/star.

Once stars begin to explode, the injection of mechanical energy is dominated by supernova explosions.
The solid curves in Fig.~\ref{fig:luminosity} represents the total (winds plus supernovae) power in the cluster, and has been computed assuming that each supernova releases $10^{51}$~erg of mechanical energy over a relaxation time of about 1 Myr.
The curves show that the power injection from supernovae stays roughly constant for few tens of Myr, which corresponds to the explosion time of the lightest stars ($M_* \lesssim 10~M_{\odot}$).
The average power of supernovae over the cluster lifetime is $\lesssim 10^{36}$~erg/s/star.
Therefore, the total power (winds plus supernovae) is $\sim 10^{36}$~erg/s, with winds contributing at the 10\% level \cite{thibault}.
 
Despite significant fluctuations, the total average power injected by stars stays remarkably constant over few tens of Myr for massive clusters (more than $\sim$100 massive stars).
Its value is indicated with dashed lines in Fig.~\ref{fig:luminosity}, and can be written as:
\begin{equation}
{\cal P}_{tot} = \langle L_w \rangle N_* \sim 3 \times 10^{51} \left( \frac{N_*}{100} \right) \rm erg/Myr
\end{equation}
and can be used to estimate the expansion rate of a bubble inflated by a star cluster.

\subsection{Expansion rate of the forward shock}

The expansion rate of the forward shock of a bubble inflated by a massive star cluster can be derived exactly as for the case of a single stellar wind, substituting in Eq.~\ref{eq:ss} $L_w$ with ${\cal P}_{tot}$.
The forward shock radius and velocity read \cite{SB,thibault}:
\begin{eqnarray}
\label{eq:Rsb}
R_s &\sim& 2.6 \times 10^2 \left( \eta \frac{N_*}{100} \right)^{1/5} \left( \frac{n_0}{{\rm cm}^{-3}} \right)^{-1/5} \left( \frac{t}{10~{\rm Myr}} \right)^{3/5} \rm pc \\
u_s &\sim& 15 \left( \eta \frac{N_*}{100} \right)^{1/5} \left( \frac{n_0}{{\rm cm}^{-3}} \right)^{-1/5} \left( \frac{t}{10~{\rm Myr}} \right)^{-2/5} \rm km/s
\label{eq:usb}
\end{eqnarray}
where $\eta$ is a correction factor that can be derived from more accurate studies (e.g. a better description of radiative losses, or of the interface between the shell and the interior, etc.
Such a parameter can be estimated thanks to numerical simulations of interstellar bubbles \cite{yadav,gupta} or, more pragmatically, from observations \cite{thibault}.
The latter method gives, with a quite large uncertainty, $\eta \approx 0.22$ \cite{thibault}.
Also the density and temperature inside the bubble follow from the same procedure used to derive Eq.~\ref{eq:Tb} and \ref{eq:nb}, and are equal to \cite{thibault}:
\begin{eqnarray}
\label{eq:TT}
T &\sim& 2.0 \times 10^6 \left( \eta \frac{N_*}{100} \right)^{8/35} \left( \frac{n_0}{{\rm cm}^{-3}} \right)^{2/35} \left( \frac{t}{10~{\rm Myr}} \right)^{-6/35} \rm K \\
\label{eq:nn}
n &\sim& 7.0 \times 10^{-3} \left( \eta \frac{N_*}{100} \right)^{6/35} \left( \frac{n_0}{{\rm cm}^{-3}} \right)^{19/35} \left( \frac{t}{10~{\rm Myr}} \right)^{-22/35} \rm cm^{-3}
\end{eqnarray}

The time evolution of the radius and velocity of the forward shock are shown in the top panel of Fig.~\ref{fig:Mach}.
As for in Fig.~\ref{fig:weaver}, curves are plotted in the range of times spanning from the end of the adiabatic phase to the beginning of the pressure-confined one (${\cal M} \sim 1$).
The radius of the bubble becomes larger than the half thickness of the Galactic disk ($\sim 100$~pc, indicated as a dashed line in the figure) before entering the pressure-confined phase.
When that happens, the bubble becomes more and more elongated in a direction perpendicular to the disk, as it is easier to expand in an ambient medium of lower density.
Eventually, the bubble breaks out in the Galactic halo, creating collimated structures called chimneys, through which matter and energy are transported to the halo \cite{SBhalo}.

\subsection{The wind termination shock: compact and loose clusters}
\label{sec:extended}

The expansion rate of the forward shock and the properties of the gas in the bubble (region {\it ii}) have been derived above following exactly the same procedure adopted for the case of a bubble inflated by a single stellar wind.
On the other hand, this cannot be done for the innermost region, i.e. that contained within the WTS (region {\it i}).
The reason for that is that star clusters are not point-like objects, and therefore the mechanical energy is injected by stellar winds in a spatially extended region.
This scenario was investigated in \cite{chevalier} and will be briefly summarised here.
 
Consider a cluster composed of $N_*$ massive stars distributed homogeneously over a spherical region of size $R_c$. 
Typical values for $R_c$ are of the order of few parsecs \cite{krumholzreview}.
For simplicity, take stars to be all identical, each blowing a wind of mass loss rate $\dot{M}_w$ and injecting mechanical energy at a rate $L_w$. 
Assume also that winds from individual stars will merge to form a collective outflow of matter (a situation where this is not the case will be described below).
Then, the total rate of injection of matter and mechanical energy are $\dot{M}_{tot} = N_* \dot{M}_w$ and ${\cal P}_{tot} = N_* L_w$.

\begin{figure}
\centering
\includegraphics[width=0.7\textwidth]{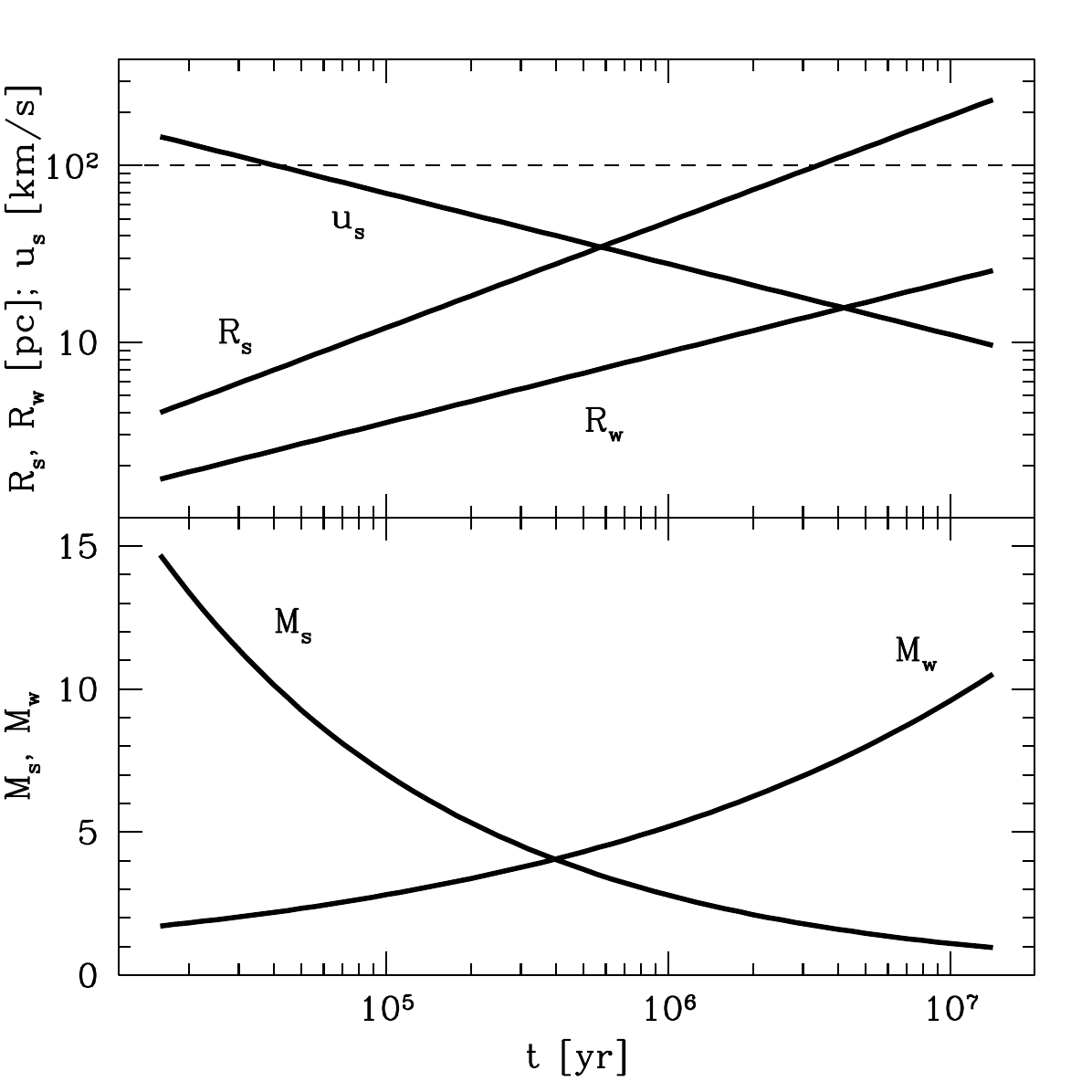}     
\caption{{\bf Top panel:} Time evolution of the forward shock radius $R_s$ and velocity $u_s$ and of the collective WTS radius $R_w$ for an interstellar bubble inflated in an ISM of density $n_0 = 1$~cm$^{-3}$ by a cluster of massive stars of total kinetic power $L_w = 10^{38}$~erg/s and terminal velocity $u_w = 3000$~km/s. {\bf Bottom panel:} time evolution of the Mach number of the forward shock and of the WTS. The latter has been computed assuming a radius of the star cluster equal to $R_c = 3$~pc. The time interval on the x-axis spans from the end of the adiabatic phase to the beginning of the pressure-confined one.}
\label{fig:Mach}
\end{figure}

For $R \gg R_c$, the cluster can indeed be considered as a point source of mass and energy, and therefore the stationary solution given by Eq.~\ref{eq:rhow} must be recovered, with the terminal velocity given by $u_w = 2 {\cal P}_{tot}/\dot{M}_{tot}$. 
This implies that the position of the WTS can be computed exactly as done in Eq.~\ref{eq:WTS}, to give:
\begin{equation}
\label{eq:CWTS}
R_w \sim 35  \left( \eta \frac{N_*}{100} \right)^{3/10} \left( \frac{n_0}{{\rm cm}^{-3}} \right)^{-3/10} \left( \frac{u_w}{3000~{\rm km/s}} \right)^{-1/2} \left( \frac{t}{10~{\rm Myr}} \right)^{2/5} \rm pc
\end{equation}
which has been plotted in the top panel of Fig.~\ref{fig:Mach}, together with the WTS velocity:
\begin{equation}
\dot{R}_w \sim 1.4  \left( \eta \frac{N_*}{100} \right)^{3/10} \left( \frac{n_0}{{\rm cm}^{-3}} \right)^{-1/5} \left( \frac{u_w}{3000~{\rm km/s}} \right)^{-1/2} \left( \frac{t}{10~{\rm Myr}} \right)^{-3/5} \rm km/s
\end{equation}

On the other hand, if energy is injected in an extended and roughly spherical region of radius $R_c$, symmetry imposes that the fluid velocity in the centre of the star cluster ($R = 0$) must vanish.
Therefore, the fluid has to accelerate from a velocity $u = 0$ in $R = 0$ to $u = u_w$ for $R \gg R_c$.
This is possible only if the gas pressure does not vanish ($P_w \ne 0$), but rather decreases towards larger radii, so that the gas is pushed outward by the $\nabla P_w$ force.
It follows that the sound speed in the wind $c_{s,w}$ is also non vanishing, and therefore the Mach number of the wind termination shock will remain finite.
It can be shown (following a somewhat lengthy calculation that can be found here \cite{chevalier}), that at large enough radii the shock Mach number scales as:
\begin{eqnarray}
\label{eq:lowmach}
{\cal M}_w &=& \frac{u_w-\dot{R}_w}{c_{s,w}} \sim \frac{u_w}{c_{s,w} } \nonumber \\
&\sim& 13  \left( \eta \frac{N_*}{100} \right)^{1/5} \left( \frac{n_0}{{\rm cm}^{-3}} \right)^{-3/10} \left( \frac{u_w}{3000~{\rm km/s}} \right)^{-1/3} \left( \frac{R_c}{3~{\rm pc}} \right)^{-2/3}\left( \frac{t}{10~{\rm Myr}} \right)^{4/15} 
\end{eqnarray}
Note that for $R_c \rightarrow 0$ the Mach number diverges, and this justifies why the WTS of an individual (point-like) star is invariably assumed to be very strong.
In fact, for an isolated star, $R_c$ would correspond to the region of wind launching, which is very small, being of the order of few stellar radii \cite{wtssonic}.
The Mach number of the WTS is shown in the bottom panel of Fig.~\ref{fig:Mach}, together with the Mach number of the forward shock.
Remarkably, they follow an opposite trend: the Mach number of the forward shock gradually decrease, while that of the WTS increases with time.
In particular, the WTS is weak (Mach number of the order of a few) for a quite long time, and, as discussed in the following, this might have an impact on particle acceleration.

To conclude, a discussion on the actual formation of the WTS is on order.
The assumption of a spatially extended injection of mechanical energy introduces a scale length into the problem, i.e. the radius of the star cluster $R_c$.
In deriving Eq.~\ref{eq:CWTS} it was implicitly assumed that the shock does form around the cluster, but a necessary condition for that to happen is $R_w > R_c$.
Star clusters can then be classified as compact when $R_c \ll R_w$ or loose in the opposite case $R_c \gg R_w$.
The formation (or non formation) of the collective WTS in the former (latter) case has been confirmed by means of hydrodynamical simulation \cite{gupta}.
In loosely bound clusters, each star may form its own, strong, WTS, and no large scale collective shock appears.

\subsection{Final remarks on interstellar bubbles inflated by star clusters}

\begin{figure}
\centering
\includegraphics[width=0.7\textwidth]{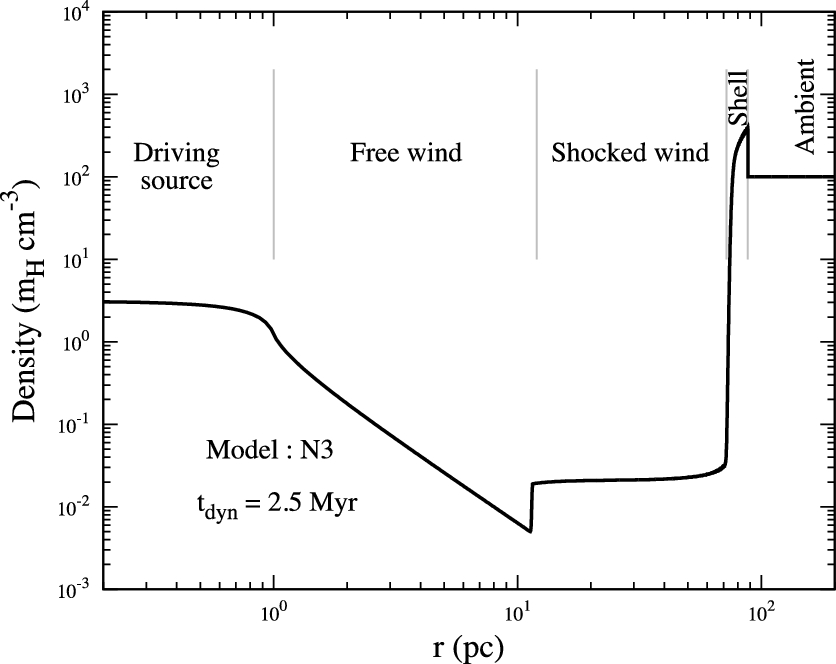}     
\caption{Density profile around a compact star cluster. Figure from \cite{gupta2}.}
\label{fig:gupta}
\end{figure}

Fig.~\ref{fig:gupta}, taken from \cite{gupta2}, shows the density profile for a compact cluster.
Energy is injected in an extended region (driving source region) having the size of the star cluster.
Such region is characterised by a mildly varying density.
Moving outwards one finds the wind region ($\varrho \propto R^{-2}$), the bubble containing the shocked wind material (roughly constant density), and the dense shell where the shocked ISM is accumulated.
As seen above, the density profile of a loose cluster will differ in the innermost region, as each star will form its own WTS, and a collective shock will not form around the cluster \cite{gupta}.

Fig.~\ref{fig:gupta} provides an appropriate description for the density profile around a star cluster during the first few megayears of its lifetime only.
After this time, supernovae will begin to explode, and this will have a dramatic impact on the density profile.
In fact, as seen in Fig.~\ref{fig:luminosity}, for rich clusters the average mechanical power injected by stellar winds and supernova explosions stays roughly constant throughout the entire cluster lifetime.
This implies that the evolution of the shell (forward shock) radius versus time does not change when supernovae overcome stellar winds as sources of energy.
In fact, numerical simulations showed that the $R_s \propto t^{3/5}$ scaling still provides a good descriptions of the evolution of interstellar bubbles even in the case of poor clusters, where only few supernovae explode (e.g. \cite{diehl}).

\begin{figure}
\centering
\includegraphics[width=0.99\textwidth]{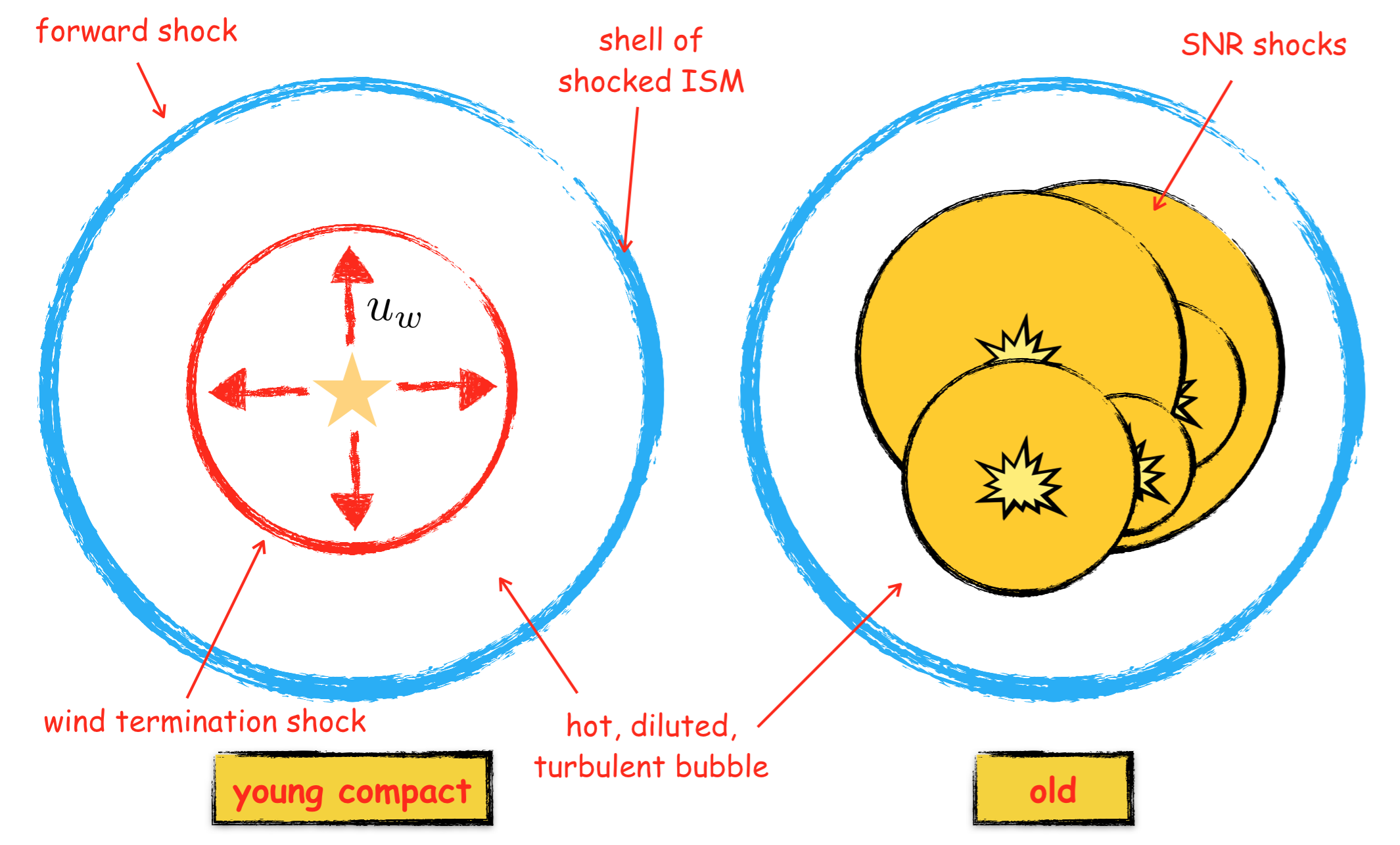}     
\caption{Sketch of the structure of a young and compact star cluster (left) and of an old one (right). Here, a cluster is called old if supernovae already began to explode.}
\label{fig:cartoon}
\end{figure}

On the other hand, the internal structure of the bubble is different before and after the onset of stellar explosions.
This is illustrated by the cartoon in Fig.~\ref{fig:cartoon}, where a sketch of the structure of a young and compact cluster is given on the left, while an older cluster is represented on the right.
As it will be discussed extensively in the following, the acceleration of particles in young clusters is likely to take place at the collective WTS (or at the individual WTSs for loose clusters), and a relatively simple (i.e., spherically symmetric, quasi-stationary) setup can be adopted to describe acceleration.
This is not the case for older clusters, where acceleration is expected to take place in the turbulent bubble, whose gas is repeatedly swept by a series of SNR shocks, possibly colliding with each other and maintaining in this way an enhanced level of turbulence.
In most cases, such systems are not expected to be spherically symmetric nor quasi-stationary, making the study of the acceleration mechanisms at work a very complicated issue.

Finally, all the results presented in this Chapter have been derived by assuming an homogeneous ISM outside of the bubble.
In fact, the ISM is a multi-phase plasma, made of cold and dense clouds surrounded by dense warm envelopes which are in turn embedded in a diffuse and hot gas that occupies most of the volume  \cite{ISM}.
The forward shock of the interstellar bubble propagates then in the diffuse phase of the ISM.
On the other hand, dense clouds can survive the passage of the forward shock and, once inside of the bubble they begin to evaporate, loading the system with mass.
It has been shown that in this case the evolution of the forward shock scales with time as $R_s \propto t^{\alpha}$, with $\alpha = 7/10$, which slightly differs from the canonical $\alpha = 3/5$ derived above \cite{threephases}.

\section{Star clusters as particle accelerators}
\label{sec:acceleration}

Three classic questions in particle acceleration in astrophysical environments are (e.g. \cite{lukerecent}):
\begin{enumerate}
\item{What is the origin of accelerate particles?}
\item{What is the origin of the energy that the particles acquire?}
\item{Where are the acceleration sites? or, equivalently: What are the acceleration mechanisms?}
\end{enumerate}

The first question deals with CR composition.
As discussed in the Introduction, some isotopic anomalies observed in the local flux of CRs require that a small but non negligible fraction of the particles which are accelerated come from Wolf-Rayet wind material \cite{22Ne}.
Data are best explained if such material is directly accelerated at the stellar WTS, and not injected in the circumstellar bubble to be then accelerated by e.g. a SNR shock \cite{vincent}.
For this reason, the acceleration of particles at WTS will be discussed in Sec.~\ref{sec:accWTS} below.

As seen in Sec.~\ref{sec:clusterwind}, the overall mechanical power of massive stars is dominated by supernova explosions, while stellar winds contribute roughly at he 10\% level.
This means that the acceleration of particles at WTS cannot provide the necessary amount of energy to explain Galactic CRs (second question in the list above).
For this reason, Sec.~\ref{sec:accSB} will be devoted to the description of the acceleration of particles in superbubbles at late times, i.e., when supernovae has already began to explode.
The acceleration mechanism is not simply diffusive acceleration at SNR shocks, but it is likely the result of the interplay of SNR shocks and plasma turbulence \cite{bykovrev,thibaultPhD}.
Understanding particle acceleration in superbubbles is extremely important.
The reason is that most stars form in clusters, and therefore the contribution to Galactic CRs from star clusters is likely to exceed that from isolated SNRs.
Somewhat surprisingly, despite this fact the standard model for CR origin relies on particle acceleration at isolated SNR shocks.

What said above also addresses question number three in the list: the particle acceleration sites in and around star clusters are most likely the WTS and the diluted region containing the shocked WTS material.
The forward shock might also accelerate particles, but its slow velocity (tens of km/s) won't allow to accelerate particles to extremely high energies  \cite{thibault,thibaulthillas}.
Moreover, as seen in Sec.~\ref{sec:snowplow}, during most of the bubble lifetime the forward shock is radiative. 
As most of the energy flowing through the shock is radiated away, it is very likely that particle acceleration will be quite ineffective.

The remainder of this Section will be devoted to an estimate of the maximum energy that accelerated particles can achieve in star clusters, and to some simplified calculations aimed at estimating the shape of the particle spectra emerging in these objects.
Remarkably, the estimate of the maximum energy can be obtained using a very simple argument based on basic electrodynamics, while particle spectra will be obtained solving partial differential equations.

\subsection{The maximum energy of accelerated particles: the Hillas criterion}
\label{sec:hillas}

All acceleration mechanisms taking place in astrophysical environments rest on the interaction between charged particles and electromagnetic fields.
In order to be accelerated, a particle carrying an electric charge $e$ must be subject to a force having a non-negligible component along the particle direction of motion, defined by its velocity $\vec{v}$.
This rules out static magnetic fields $\vec{B}$ as particle accelerators, as they exert a force $\vec{F} = (e/c) \vec{v} \times \vec{B}$ orthogonal to the velocity of the particle.
On the other hand, a static electric field $\vec{\cal E}$ will accelerate a charged particle via the electrostatic force $\vec{F} = e \vec{\cal E}$.

Consider now a region of space of size $L$ where a uniform electric field is present. 
A particle crossing the region will gain an energy:
\begin{equation}
\label{eq:Estatic}
\Delta E = e {\vec{\cal E}} L
\end{equation}
which can be very large if an intense electric field occupies a large region of space.
Unfortunately, astrophysical plasmas are characterised by very large values of the electric conductivity.
This means that any charge excess in a plasma (let's say of charge density $\varrho_e$) will be rapidly neutralised by the motion of charges of opposite sign in the plasma, making it very difficult to maintain a static, strong, and large scale electric field, as $\nabla \vec{\cal E} = 4 \pi \varrho_e \sim 0$.

In turbulent plasmas, time varying magnetic fields induce electric fields, as stated by Faraday's law:
\begin{equation}
\nabla \times \vec{\cal E} = - \frac{1}{c} \frac{\partial \vec{B}}{\partial t}
\end{equation}
To obtain an order of magnitude estimate of the intensity of the induced electric fields, the equation above can be simplified by setting $\nabla \times \rightarrow 1/L$ and $\partial / \partial t \rightarrow T$, where $L$ and $T$ are the characteristic length and time scales over which electromagnetic fields vary.
Introducing also the characteristic velocity of motions in the plasma, which has to be of the order $U = L/T$, one gets ${\cal E} \sim (U/c) B$.
Substituting into Eq.~\ref{eq:Estatic} and setting $\Delta E = E_{max}$ gives:
\begin{equation}
\label{eq:hillas}
E_{max} \sim \left( \frac{e}{c} \right) B U L 
\end{equation}
which is universally known as the {\it Hillas criterion} \cite{hillas} and represents the maximum energy that a particle can attain in an accelerator of size $L$, characterised by plasma motions of velocity $U$, and containing a magnetised plasma of magnetic field strength $B$.
The implicit assumption done to derive the Hillas criterion is that particles do not suffer energy losses, and therefore the value of $E_{max}$ has to be considered the most optimistic one (for a treatment of energy losses in this context see \cite{derishev}).

The Hillas criterion is widely used because of its predictive power and its simplicity. 
It provides an estimate of the maximum particle energy allowed by electrodynamics, without the need to specify the nature of the acceleration mechanism!
Unfortunately, while the size $L$ and the characteristic plasma velocity $U$ can be measured for a large number of astrophysical objects, the magnetic field strength $B$ is very often unknown as it is difficult to constrain it from observations \cite{Bfields}.
It is therefore convenient to rewrite Eq.~\ref{eq:hillas} as:
\begin{equation}
B \sim 3 \times 10^2 \left( \frac{E_{max}}{\rm PeV} \right) \left( \frac{U}{1000~{\rm km/s}} \right)^{-1} \left( \frac{L}{\rm pc} \right)^{-1} \mu {\rm G}
\end{equation}
which defines the minimum magnetic field strength necessary to accelerate CR {\it protons} up to an energy $E_{max}$.

The expression above can be applied, for example, to the collective WTS of a very compact (point like) star cluster.
In this case, the characteristic length would be the radius of the WTS, while the characteristic plasma velocity would be the wind terminal velocity. 
Setting (see Eq.~\ref{eq:CWTS} and/or Fig.~\ref{fig:Mach}) $U = u_w \approx 3000$~km/s and $L = R_w \approx 10$~pc one gets that, in order to accelerate protons up to the energy of the CR knee (about 4 PeV), the magnetic field strength should be at least of the order of $B \approx 40  ~\mu$G.
Such a value of the magnetic field corresponds to a magnetic pressure of $P_m = B^2/8 \pi$, which can be compared to the shock ram pressure $P_r = \varrho_w u_w^2$.
Making use of Eqns.~\ref{eq:rhow} and \ref{eq:hillas}, the ratio between these two pressures reads:
\begin{equation}
\frac{P_m}{P_r} = \frac{1}{4} \left( \frac{c}{e} \right)^2 \frac{E_{max}}{{\cal P}_{tot} u_w} \sim 1.3 \left( \frac{E_{max}}{4~{\rm PeV}} \right)^2 \left( \frac{{\cal P}_{tot}}{3 \times 10^{51} {\rm erg/Myr}} \right)^{-1} \left( \frac{u_w}{3000~{\rm km/s}} \right)^{-1}
\end{equation}
In order to conserve energy, the magnetic pressure should not exceed the ram pressure, and in fact a realistic condition would read $P_m/P_r \ll 1$.
This implies that acceleration at the WTS up to the particle energies that characterise the knee is possible only for very powerful clusters, having mechanical luminosities significantly exceeding $\sim 3 \times 10^{51}$~erg/Myr $\sim 10^{38}$~erg/s.
Remarkably, the very same result was obtained from a sophisticated study of particle acceleration at the WTS \cite{morlino}, and this demonstrates that the Hillas criterion is a very powerful tool.

The Hillas criterion can also be used to constrain the maximum energy of particles accelerated in the turbulent and rarefied interstellar bubble \cite{thibaulthillas}.
In this case, the size of the accelerator can be taken to be equal to the radius of the bubble, $L = R_s$.
The value of the parameter $U$ may be taken to be equal to the velocity $u_t$ of turbulent motions inside the bubble.
The energy density of the turbulent gas is $\varrho u_t^2$, where $\varrho$ is the gas density inside the bubble, while that of the magnetic field is $B^2/8 \pi$.
To conserve energy, both these energy densities will have to be at most of the order of the thermal energy density $(3/2) n k T$, as it was estimated from Eqns.~\ref{eq:TT} and \ref{eq:nn}.
From this conditions, and making use of Eq.~\ref{eq:hillas}, an upper limit on the maximum proton energy that can be achieved in a superbubble can be derived.
It reads:
\begin{equation}
E_{max} \ll 1 ~ \left( \eta \frac{{\cal P}_{tot}}{3 \times 10^{51}{\rm erg/Myr}} \right)^{18/35} \left( \frac{n_0}{\rm cm^{-3}} \right)^{9/70} \left( \frac{t}{10~{\rm Myr}} \right)^{4/35} ~ \rm PeV
\end{equation}
and shows that it is highly unlikely that turbulent superbubbles are able to accelerate protons beyond PeV energies.

\subsection{Particle acceleration at the wind termination shock}
\label{sec:accWTS}

The spectrum of energetic particles accelerated at a spherical WTS can be derived solving the transport equation for CRs, first derived in \cite{parker} (see also Blasi's lecture notes, this volume).
The transport equation describes the evolution in time $t$ of the isotropic part of the particle distribution function $f(t,p,\vec{R})$, which is also a function of the particle momentum $p$ and of the spatial coordinate $\vec{R}$. In this notation, the number density of energetic particles at a given time and place is $n = 4 \pi \int {\rm d}p~ p^2 f$.
The steady state (time independent) solution of the problem is obtained solving the equation:
\begin{equation}
\label{eq:parker}
u \frac{\partial f}{\partial R} = \frac{1}{R^2} \frac{\partial}{\partial R} \left( R^2 D \frac{\partial f}{\partial R} \right) + \frac{p}{3 R^2} \frac{{\rm d} \left( u R^2 \right)}{{\rm d} R} \frac{\partial f}{\partial p} 
\end{equation}
where spherical symmetry has been assumed.
Here, $u(R)$ represents the velocity profile of the gas and $D(R,p)$ the diffusion coefficient of particles of momentum $p$.
The term on the left hand side describes the advection of particles with the flow, while the two terms on the right hand side account for energetic particles spatial diffusion in the turbulent ambient magnetic field and particle acceleration/deceleration induced by fluid compression/decompression.
Radiative energy losses are ignored (and for CR protons this is very often a safe assumption).

In general, Eq.~\ref{eq:parker} is solved numerically (e.g. through a finite differences scheme), as an exact analytic solution is known only for the (quite unphysical, unfortunately) case of a diffusion coefficient which is independent on particle momentum \cite{WTStheory}.
However, approximate analytic solutions can still be obtained in the limit of both large and small particle momenta.
This can be seen by comparing the advection and diffusion terms in the equation, i.e., the terms depending on the spatial variation of CRs in the system.
In general, $u$, $D$, and $f$ may all vary with position.
However, in order to obtain an order of magnitude estimate  the following substitutions can be made:
\begin{eqnarray}
u(R) &\rightarrow& U \\
D(R) &\rightarrow& \kappa \\
\frac{\partial}{\partial R} &\rightarrow& \frac{1}{L}
\end{eqnarray} 
where $U$, $\kappa$, and $L$ represent some characteristic values for the fluid velocity, the diffusion coefficient, and the spatial scale over which significant variations of the various physical quantities occur, respectively.
Once these substitutions are applied, the ratio between the advection and the diffusion term in Eq.~\ref{eq:parker} is \cite{florinski}:
\begin{equation}
\label{eq:lowhigh}
\frac{u \frac{\partial f}{\partial R}}{\frac{1}{R^2} \frac{\partial}{\partial R} \left( R^2 D \frac{\partial f}{\partial R} \right)} \longrightarrow \frac{U L}{\kappa}
\end{equation}
As the CR diffusion coefficient increases with particle momentum (see Blasi's lecture, this volume), a low and high energy regime can be defined according to the conditions $U L/\kappa \gg 1$ and $U L/\kappa \ll 1$, respectively.
In the low energy regime, then, advection dominates over diffusion, while the opposite is true in the high energy domain.
Approximate analytic solutions have been derived for both the low \cite{luke} and high \cite{fisk} energy limits.

Which are the appropriate values for the physical quantities $U$, $\kappa$, and $L$?
For definiteness of discussion, consider a setup of the problem where the diffusion coefficient downstream of the WTS is so small that for all practical purposes its value can be considered to be very close to 0.
Such an extreme assumption can be justified by recalling that the magnetised plasma downstream of a shock is expected to be highly turbulent \cite{giacalone}, and that in an highly turbulent medium particles are scattered very effectively and therefore diffusion is strongly suppressed.
It follows that accelerated particles downstream of the shock will simply follow the fluid flow and be advected outwards up to the edge of the bubble, at $R_s$, where they will freely escape in the ISM (as the diffusion coefficient there is much larger).
If a velocity profile scaling as $u \propto R^{-2}$ is adopted in the WTS downstream region (as done, e.g., in \cite{luke} and \cite{morlino}) the transport equation (Eq.~\ref{eq:parker}) reduces to a description of pure advection, $u ~{\rm d}f/{\rm d}R=0$. 
The CR particle distribution function is then spatially homogeneous within the bubble ($R_w < R < R_s$) and equal to $f_w \equiv f(R = R_w)$, regardless of the position of the forward shock $R_s$.
It follows that $L = R_s$ is not a good choice, as $R_s$ does not influence at all $f_w$.
The other two spatial scales in the problem are the radius of the star cluster $R_c$ and that of the WTS $R_w$.
However, if one makes the further simplifying assumption that mechanical energy is injected in a very small (almost pointlike, i.e. $R_c \ll R_w$) region, then the only possible choice is to set $L = R_w$.
Assuming a point-like source of energy injection also implies that the wind velocity is constant for any $R < R_w$ (see Sec.~\ref{sec:extended}) and therefore $U = u_w$.

In order to chose the value of $\kappa$, notice that the problem simplifies significantly under the assumption that $D \rightarrow 0$ as $R \rightarrow 0$ \cite{luke}.
If this is the case, outward advection dominates close to $R = 0$, implying that the boundary condition for the CR particle distribution function must be $f(R = 0) = 0$.
At this point, to ease computations, a linear scaling of the diffusion coefficient with the radial coordinate is often assumed (see e.g. \cite{florinski} or \cite{fisk}):
\begin{equation}
\label{eq:diff}
D(R,p) = D_w(p) \left( \frac{R}{R_w} \right)
\end{equation}
where $D_w$ is the CR diffusion coefficient immediately upstream of the WTS.
After introducing this parameterisation, it seems convenient to set $\kappa = D_w$, so that the boundary between low and high energy regime is set by the condition $u_w R_w/D_w= 1$ (see Eq.~\ref{eq:lowhigh}).

\subsubsection{The low energy limit}

The low energy limit is defined by the condition $u_w R_w/D_w \gg 1$, which can be rewritten as:
\begin{equation}
\label{eq:difflength}
l_d = \frac{D_w}{u_w} \ll R_w 
\end{equation}
where $l_d$ is called diffusion length.
Ignoring for a moment the spatial dependence of $D_w$, the quantity $l_d$ represents the diffusion length of particles ahead (upstream) of the shock.
In other words, accelerated particles are not able to reach distances from the shock exceeding significantly $l_d$, as in that case outwards advection dominates over spatial diffusion.
This can be easily proven by recalling that in a time $t$ advection would displace particles by an amount $l_a = u_w t$, while diffusion would spread particles over a region of size $l_d \sim \sqrt{D_w t}$.
The two displacements are equal for a characteristic time $t_d$, which gives $l_a \equiv l_d = D_w/u_w$.
For times longer than $t_d$ advection would dominate over diffusion and keep accelerated particles within a diffusion length from the shock surface.
Thus, the condition expressed by Eq.~\ref{eq:difflength} means that particle acceleration happens in a region upstream of the shock whose extension is much smaller than the WTS radius.
Therefore, the sphericity of the shock can be ignored when studying CR acceleration at low enough particle energies.

\begin{figure}
\centering
\includegraphics[width=0.7\textwidth]{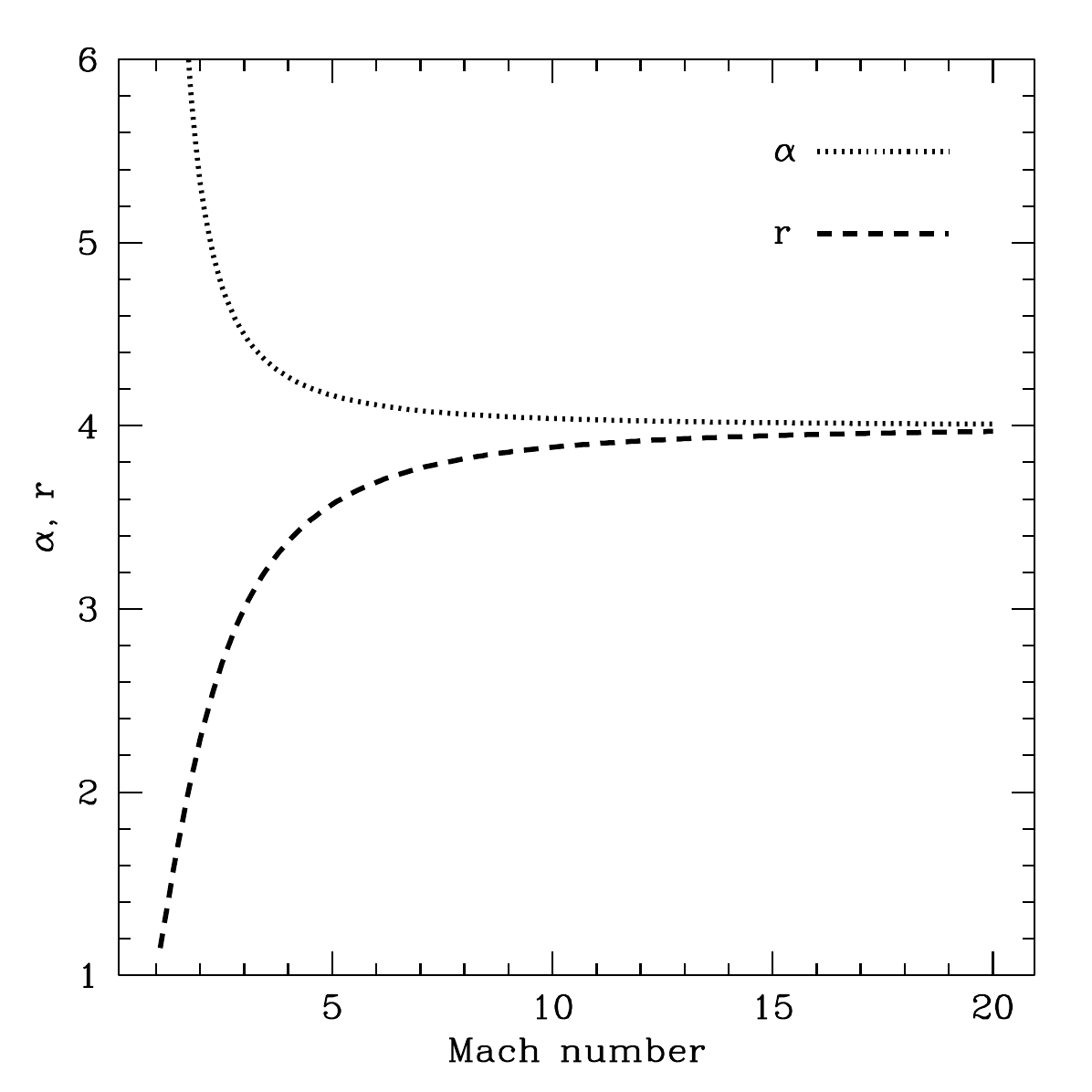}     
\caption{Shock compression factor (dashed line) and spectral slope of the accelerated particles (dotted line) as a function of the shock Mach number (see Eq.~\ref{eq:slope}).}
\label{fig:slope}
\end{figure}

Diffusive acceleration at plane shocks has been discussed by Caprioli (this volume).
The spectrum of particles accelerated at a plane shock can be obtained by solving the CR transport equation (the analogue of Eq.~\ref{eq:parker} in one dimension and cartesian coordinates).
It is a power law in particle momentum $f(p) \propto p^{-\alpha}$ where the slope $\alpha$ depends on the shock Mach number ${\cal M}$ or by the shock compression factor $r$ as:
\begin{equation}
\label{eq:slope}
\alpha = \frac{3~r}{r-1} = \frac{4~{\cal M}^2}{{\cal M}^2 - 1} \, .
\end{equation}
These dependences are shown in Fig.~\ref{fig:slope}.

It is interesting to remark that, as seen in Sec.~\ref{sec:extended}, the WTS for a compact star cluster is not very large (see Eq.~\ref{eq:lowmach} and Fig.~\ref{fig:Mach}), and therefore the spectrum of accelerated particles is expected to be slightly steeper than 4.
For example, a slope $\alpha = 4.1$ (4.4) would correspond to a Mach number ${\cal M} = 6.4$ (3.3).
Slopes slightly larger than 4 are those needed to explain Galactic cosmic rays (see Introduction), but one should remember that WTSs can only provide a minor contribution to the observed intensity of CRs.
Therefore such agreement between predictions and expectations should probably be considered as a coincidence.

\subsubsection{The high energy limit}

In the high energy limit the advection term can be neglected, as $u_w R_w/D_w \ll 1$, and the transport equation reduces to:
\begin{equation}
\label{eq:parkerhigh}
\frac{1}{R^2} \frac{\partial}{\partial R} \left( R^2 D \frac{\partial f}{\partial R} \right) + \frac{p}{3 R^2} \frac{{\rm d} \left( u R^2 \right)}{{\rm d} R} \frac{\partial f}{\partial p} = 0
\end{equation}
Integrating between $R_w^- = R_w-\epsilon$ and $R_w^+ = R_w + \epsilon$ where  $\epsilon$ is arbitrarily small one gets:
\begin{equation}
\label{eq:parkerhighw}
\left[ R^2 D \frac{\partial f}{\partial R} \right]_{R_w^-} + R_w^2 u_w \frac{r- 1}{3~ r} p \frac{\partial f_w}{\partial p}= 0
\end{equation}
where we used $D(R_w^+) = 0$ and the fact that the fluid velocity immediately upstream (downstream) of the shock is $u_w$ ($u_w/r$), $r$ being the shock compression factor.	

The solution of Eq.~\ref{eq:parkerhigh} is obtained by setting $f(R,p) = f_w(p) f_r(R)$ and noticing that combining Eqns.~\ref{eq:parkerhigh} and \ref{eq:diff} gives $f_r(R) = (R/R_w)^{\gamma}$, where $\gamma$ 	will be determined later.
Eqns.~\ref{eq:parkerhigh} and \ref{eq:parkerhighw} can now be rewritten as:
\begin{equation}
\frac{p}{f_w} \frac{\partial f_w}{\partial p} = - \frac{3}{2} \gamma (2 + \gamma) \left( \frac{D_w}{u_w R_w} \right)
\end{equation}
and
\begin{equation}
\frac{p}{f_w} \frac{\partial f_w}{\partial p} = - \alpha \gamma \left( \frac{D_w}{u_w R_w} \right) \, ,
\end{equation}
respectively. Combining them one gets $\gamma = (2 \alpha - 6)/3$.
Finally, if CR diffusion proceeds at the Bohm rate, $D_w \propto p$, a simple integration gives the high energy behaviour of the spectrum of particles accelerated at the WTS:
\begin{equation}
f_w(p) \propto \exp \left[ - \frac{2 \alpha -6}{3} \alpha \left( \frac{D_w}{u_w R_w} \right) \right]
\end{equation}
This asymptotic solution indicates that the CR spectrum is exponentially suppressed at large energies.

A very rough description of the CR spectrum at the WTS in the entire energy domain can be obtained combining the low and high energy asymptotic solutions\footnote{This solution is not very accurate for particle energies marking the transition between a power law and an exponential cutoff spectral behaviour. A numerical solution of the problem can be found in \cite{florinski}, showing that small bumps may appear in the spectrum just before the cutoff.}:
\begin{equation}
f_w(p) \propto p^{-\alpha} \exp \left[ - \frac{2 \alpha -6}{3} \alpha \left( \frac{D_w}{u_w R_w} \right) \right] \, .
\end{equation}
Note that, expressing the exponential cutoff in terms of the particle energy, $f_w \propto \exp[-E/E_{max}]$, and making use of the definition of Bohm diffusion:
\begin{equation}
D_w = \frac{1}{3} R_L c 
\end{equation}
and of the Larmor radius of a proton of charge $e$ gyrating around a magnetic field of strength $B_w$:
\begin{equation}
R_L = \frac{pc}{eB_w}
\end{equation}
one gets:
\begin{equation}
E_{max} = \frac{9}{2 (\alpha -3) \alpha}  \left( \frac{e}{c} \right) u_w R_w B_w
\end{equation}
which is equivalent to the Hillas criterion derived in Sec.~\ref{sec:hillas} (see Eq.~\ref{eq:hillas}).
For values of the spectral slope in the range $\alpha = 4 ... 5$ the function $9/[2 (\alpha - 3) \alpha ]$ varies from $\sim$~0.45 to $\sim$~1.1.

\subsection{Particle acceleration in superbubbles}
\label{sec:accSB}

Studying the acceleration of particles in turbulent superbubbles is a very difficult task.
Acceleration of CRs may take place at WTSs and at SNR shocks.
Occasionally, shock-shock collisions may happen.
Pre-existing CRs can be reaccelerated due to second order Fermi acceleration in the highly turbulent environment.
The level of turbulence might differ in the core of the bubble, where mechanical energy is injected, and in its outskirts.
Accelerated particles can diffusively escape from the bubble, or can be advected into the halo when bubbles break out in the halo and form chimneys.
A description of sophisticated theoretical models attempting to tackle this very complex problem goes beyond the scope of this Chapter, and the interested reader is referred to the following publications.
Models for CR acceleration at multiple shock waves can be found here \cite{multipleDSA}, while models including (or trying to include) all the other physical ingredients mentioned above can be found here \cite{bykovrev,thibaultPhD,SBtheoryBykov,etienne1,thibault}.
Unfortunately, testing these models is not trivial, as observations of superbubbles are quite sparse.

\begin{figure}
\centering
\includegraphics[width=0.99\textwidth]{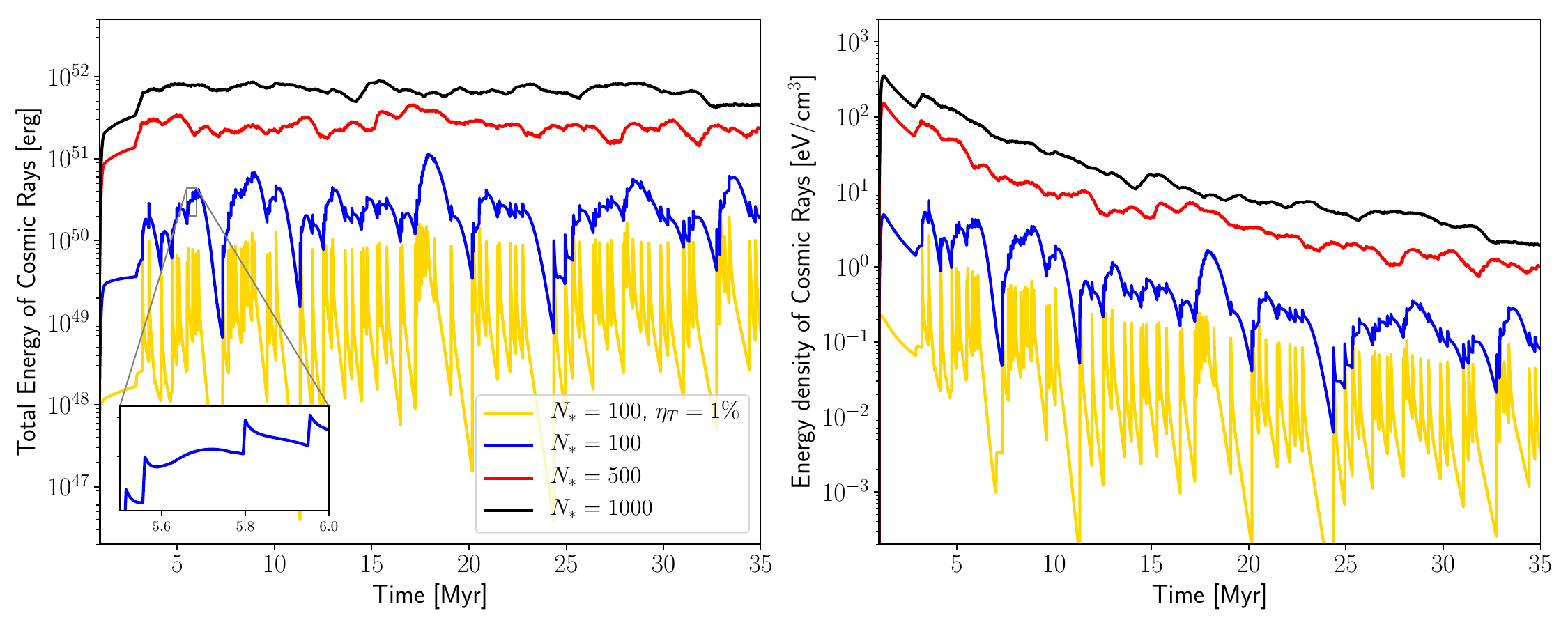}     
\caption{Simulated time evolution of the total CR energy (left) and CR energy density (right) inside a superbubble inflated by a star cluster containing initially 100, 500 and 1000 massive stars. All
the results are obtained assuming that $\eta_T =$~30\% of the mechanical energy injected in the system is converted into turbulent motions, except those shown by the yellow curves for which such efficiency is equal to 1\%. Figure from \cite{thibault}, where more details about the modelling can be found.}
\label{fig:intermittency}
\end{figure}

Probably, the two most relevant signatures of particle acceleration in superbubbles are intermittency and structured particle spectra (contrary to the featureless power laws expected when diffusive shock acceleration operates).
Intermittency is a consequence of the fact that supernova explosions are the main source of mechanical energy in a superbubble.
Assuming that all stars in a cluster are born together at time $t = 0$, the last supernova will explode in the cluster at a time equal to the lifetime of a star of $\approx 10~M_{\odot}$, i.e. $\tau_* \approx 35$~Myr (see Fig.~\ref{fig:limongi}).
If the cluster contains $N_*$ massive star that will end their life as supernovae, then a very rough estimate of the typical time between two consecutive explosions is $\Delta \tau_* \approx \tau_*/N_*$.
This can be compared with the CR diffusive escape time from the bubble, which is $\tau_{esc} \approx R_s^2/D$, where $R_s$ is the radius of the forward shock and $D$ the energy dependent CR diffusion coefficient.
If $\tau_{esc} \ll \Delta \tau_*$ CRs will be able to escape the system before the shock generated by the next supernova will inject new energetic particles.
Therefore, the bubble will empty of CRs between explosions, and this intermittent behaviour will be also reflected in the emission (for example in gamma rays) resulting from the interactions between the accelerated particles and the ambient gas.
Remarkably, this might explain why some superbubbles have been detected in gamma rays and some others not, despite their similarity (see discussion and references in \cite{thibault}).

The total CR energy stored in a bubble as a function of its age is shown in Fig.~\ref{fig:intermittency}.
The left panel refers to the total energy, while the left one to the energy density.
The latter decreases with time as the bubble volume increases.
Notice that, while for very rich clusters, hosting more than 100 massive stars, the total CR energy stays constant, for smaller clusters large fluctuations appears.
This is indeed expected, as a smaller number of stars implies a longer time between consecutive explosions, $\Delta \tau_*$.
Moreover, fluctuations are more pronounced if the bubble is less turbulent.
This can be seen by comparing the blue and yellow lines in Fig.~\ref{fig:intermittency}, which have been computed assuming that the fraction of the mechanical energy injected in the system that is converted into turbulent motions is $\eta_T =$~30\% and 1\%, respectively.
This is a consequence of the fact that CR particles are confined more effectively (i.e. their diffusion coefficient $D$ is smaller) if the level of turbulence is large.
A large diffusion coefficient corresponds to a short escape time from the system and therefore implies more intermittency.

\begin{figure}
\centering
\includegraphics[width=0.8\textwidth]{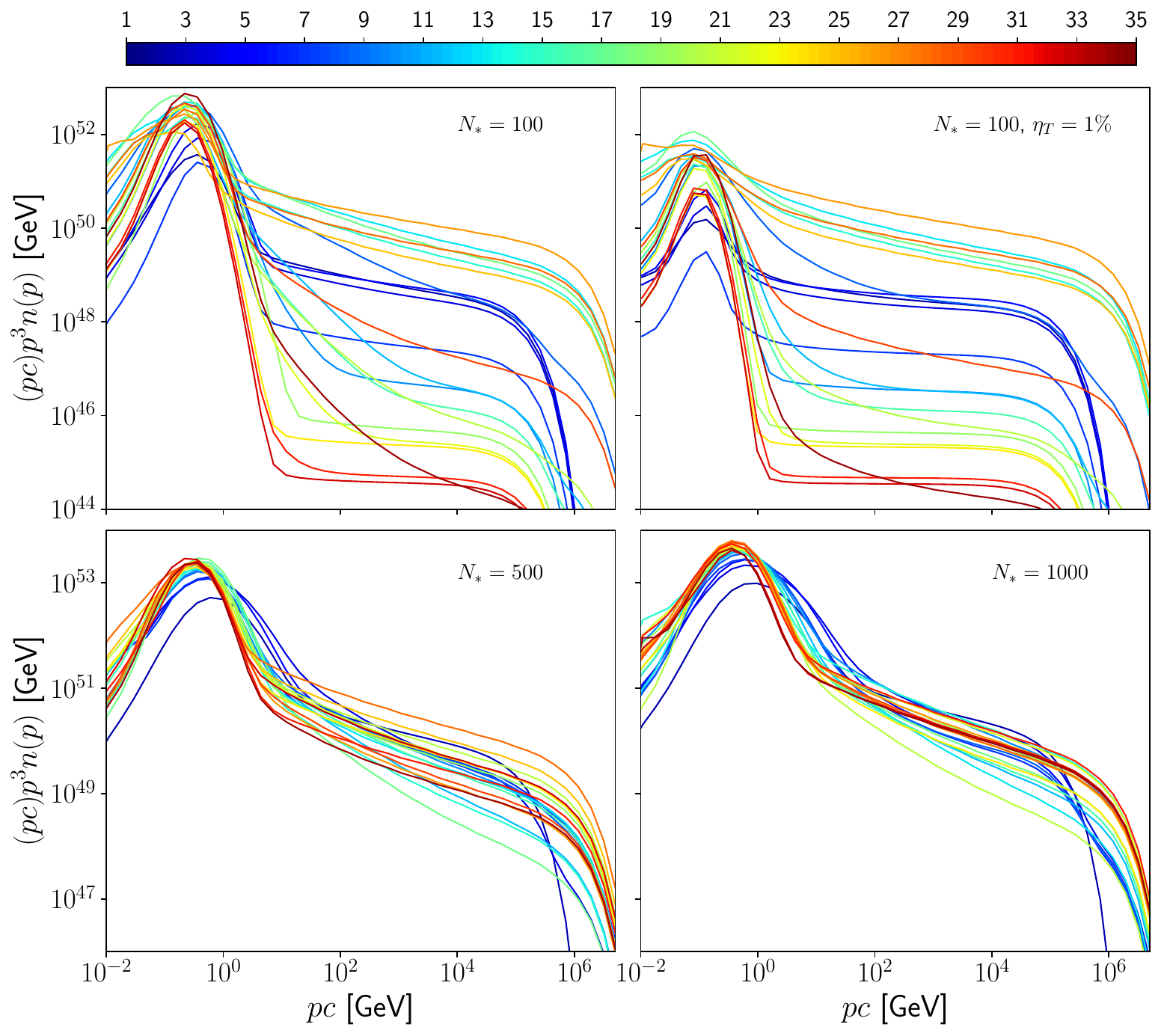}     
\caption{Spectra of cosmic rays inside superbubbles of different ages (the color scale refers to the age of the system in Myr). The number of massive stars in the cluster $N_*$ is indicated in each panel.  The parameter $\eta_T$ is defined as in Fig.~\ref{fig:intermittency}. Figure from \cite{thibault}, where more details can be found.}
\label{fig:spectra}
\end{figure}

As the diffusion coefficient is an energy dependent quantity, also the level of intermittency will depend on particle energy.
This is illustrated in Fig.~\ref{fig:spectra}, where the spectra of CRs contained within a superbubble are plotted for different times, different number of massive stars in the cluster, and different levels of turbulence (see figure caption).
The figure shows that the amount of lower energy particles stored in a superbubble does not fluctuate much.
Also in this case, the reason is that low energy CRs are characterised by a smaller diffusion coefficient $D$, and are better confined inside bubbles.
On the contrary, very large fluctuations in time are observed at large particle energies (large diffusion coefficients).

Another important result emerging from Fig.~\ref{fig:spectra} is that particle spectra are very structured, and do not resemble at all the featureless power laws which are a signature of diffusive shock acceleration.
In fact, this is due to the fact that the acceleration proceeds in a different way depending on the energy of the particles.
At low energies, second order Fermi turbulent reacceleration and Coulomb energy losses dominate, and a very pronounced bump appears in the spectrum at trans-relativistic energies.
On the contrary, at high energies particles are loss free, and the spectral shape is determined by an interplay of diffusive acceleration at SNR and WTS and diffusive escape from the system.
To conclude, a large variety of spectra could be produced inside superbubbles, and this constitutes the most important prediction to be tested with future observations of these objects.

\section{Open problems and conclusions}
\label{sec:conclusions}

The need to explain anomalies in the composition of CRs (especially the excess in the $^{22}$Ne/$^{20}$Ne ratio \cite{22Ne}) led to the suggestion that WTS of Wolf-Rayet stars might act as powerful particle accelerators \cite{casse, cesarsky}.
However, it was immediately recognised that stellar winds could provide only a fraction of the mechanical energy needed to explain the bulk of Galactic CRs, and such early estimates have been confirmed by recent studies \cite{seo}.
Then, in order to explain both the bulk of CRs and the isotopic anomalies, a scenario emerged where (at least) two classes of sources accelerate the CRs observed locally.
Supernovae explosions provide the bulk of the energy \cite{blasireview,lukerecent,myreview}, with WTS adding a small but non-negligible contribution (e.g. \cite{vincent}).

Massive stars, then, may provide the energy of {\it all} Galactic CRs.
As massive stars are rarely isolated, star clusters become natural candidate sources of CRs.
The interest towards this class of objects was recently revived by the detection of gamma-ray emission from a number of them, or from their immediate vicinity \cite{felixmassivestars,clustersgamma}.

Particle acceleration in star cluster is likely to proceed in a different way for young and old systems.
In clusters younger than few million years, stellar winds are the main source of mechanical energy, and diffusive acceleration at the WTS will most likely produce power law spectra of CRs.
For most massive clusters, the acceleration mechanism might be fast enough to accelerate protons up to the PeV domain (e.g. \cite{morlino}).
On the other hand, in older clusters the main input of energy is provided by supernova explosions.
In this case, the acceleration mechanism is not well understood, and is probably defined by an interplay between diffusive shock acceleration and reacceleration of particles in the turbulent plasma that fills the bubble \cite{bykovrev}.

The main difficulty in testing acceleration models in star cluster was connected to the scarcity of high energy observations of these objects.
However, the number of detection in gamma rays has increased steadily in the past few years, and the advent of multi-TeV detectors of unprecedented sensitivity such as LHAASO \cite{lhaaso} promise to radically impact on this field of research, especially for what concerns the search of CR PeVatrons.

On the theoretical side, the most pressing issue is the understanding of the acceleration mechanism operating in superbubbles.
To do so, a better knowledge of the plasma flow and of the magnetic field strength and structure is mandatory.
In fact, recent simulations show that also the simplest case of young (no supernova explosions) and compact star clusters blowing a wind requires detailed studies as such systems are far from the idealised spherically symmetric setup that is often assumed \cite{bykovnew}.

A solid understanding of the acceleration mechanism is also necessary in order to produce reliable predictions on the contribution of star clusters to the flux of Galactic CRs, and to estimate their impact on CR composition.
With this respect, very recent results indicate that SNR shocks expanding in the collective wind around a compact star cluster might accelerate particles well beyond PeV energy, making SNRs inside star clusters potential sources of CRs up to the transition to extragalactic CRs \cite{thibaultnew}.

Finally, the fact that the Solar system is located within a superbubble (the local bubble \cite{localbubble}) inflated by a star cluster formed about 14 million years ago \cite{bubbleage} has very important implications.
The transport of CRs in the very local ISM might be significantly affected by the magnetic field topology shaped by the inflation of the bubble, especially for low energy particles \cite{localbubblescreen,gabicilow}.
The low ambient gas density inside the local bubble might also induce effects on the production of CR secondaries \cite{LiBeBLB}.
Finally, the presence of nearby (in both time and space) massive stars and supernova explosions \cite{dieternature} must be taken into account when interpreting local CR data.
Our entire view of CRs may be biased by our location inside a superbubble.

\acknowledgments
The author acknowledges the organisers of the school (especially Carmelo Evoli) for their invitation and  Thibault Vieu, Vincent Tatischeff, and Lioni-Moana Bourguinat for discussions about cosmic rays in star clusters. He also acknowledges support from Agence Nationale de la Recherche (project CRitiLISM, ANR-21-CE31-0028).

\end{document}